\documentclass[sn-mathphys]{sn-jnl}

\usepackage{graphicx}%
\usepackage{multirow}%
\usepackage{amsmath,amssymb,amsfonts}%
\usepackage{amsthm}%
\usepackage{mathrsfs}%
\usepackage[title]{appendix}%
\usepackage{xcolor}%
\usepackage{textcomp}%
\usepackage{manyfoot}%
\usepackage{booktabs}%
\usepackage{algorithm}%
\usepackage{algorithmicx}%
\usepackage{algpseudocode}%
\usepackage{listings}%
\usepackage{comment}
\usepackage{siunitx}
\usepackage{subcaption}
\usepackage{cancel}
\usepackage{float}

\usepackage[backend=biber,style=numeric-comp,sorting=none]{biblatex}
\begin{document}

\title[Article Title]{Intrinsic and Tunable Superconducting Diode Effect in Quantum Spin Hall Systems}


\author*[1,2]{\fnm{Samuele} \sur{Fracassi}}\email{samuele.fracassi@edu.unige.it}

\author[1,2]{\fnm{Simone} \sur{Traverso}}

\author[3]{\fnm{Stefan} \sur{Heun}}

\author[1,2]{\fnm{Maura} \sur{Sassetti}}

\author[2]{\fnm{Matteo} \sur{Carrega}}

\author[1,2]{\fnm{Niccolo} \sur{Traverso Ziani}}

\affil[1]{\orgdiv{Dipartimento di Fisica}, \orgname{Università di Genova}, \orgaddress{\street{Via Dodecaneso 33}, \city{Genova}, \postcode{16146}, \state{Italy}}}

\affil[2]{\orgdiv{CNR-SPIN}, \orgaddress{\street{Via Dodecaneso 33}, \city{Genova}, \postcode{16146}, \state{Italy}}}

\affil[3]{\orgdiv{NEST}, \orgname{Istituto Nanoscienze-CNR and Scuola Normale Superiore}, \orgaddress{\street{Piazza San Silvestro 12}, \city{Pisa}, \postcode{56127}, \state{Italy}}}


\abstract{Nonreciprocal dissipationless transport has long been sought for applications in superconducting technologies. Recently, it has been implemented by the so called superconducting diode effect. Such  effect arises from an imbalance in critical supercurrents flowing in opposite directions. In this work, we theoretically demonstrate how the superconducting diode effect emerges in the quantum spin Hall phase when brought into full proximity with a superconductor. We explore two regimes: large and narrow quantum wells. In the former geometry, we show that the superconducting diode effect can be externally controlled using both magnetic and electric fields, achieving unit efficiency. In the latter regime, where tunneling between opposite edges may occur, we propose a mechanism for an intrinsic superconducting diode effect driven by edge reconstruction, which does not require external magnetic fields.}




\maketitle

\section*{Introduction}\label{sec1}
The superconducting diode effect (SDE) refers to dissipationless nonreciprocal transport in superconducting structures \cite{Hu2007,Wakatsuki2017}. It manifests in the difference in magnitude of critical currents flowing in opposite spatial directions. Since its first realization in 2020 \cite{Ando2020}, it has rapidly emerged as a central topic within the scientific community \cite{Hoshino2018,Yasuda2019,Misaki2021,Daido2022,Tanaka2022,Lu2023,Cayao2024,Reinhardt2024,Souto2024}. One reason for this success is that the SDE opens the door to engineering superconducting circuits with novel functionalities \cite{Nadeem2023}, potentially mirroring the success of its non-superconducting counterpart. Beyond its technological potential, the SDE is also of fundamental interest: since the breaking of both spatial inversion and time-reversal symmetries are necessary ingredients for its observation \cite{Ando2020}, a complete understanding of this phenomenon would deepen our knowledge of the role of symmetries in superconductivity. On the experimental side, the SDE is ubiquitous. It has been experimentally observed in a wide variety of systems, ranging from heterostructures with bulk superconductors \cite{Ando2020,YuanFu2022,Bauriedl2022,Strambini2022,Wu2022,Yun2023,Sundaresh2023,Hou2023} to Josephson junctions (often referred to as Josephson diode effect) \cite{Turini2022,Davydova2022,Souto2022,Pillet2023,Wang2024,Chieppa2025}. Equally substantial is the number of theoretical proposals and discussions surrounding the mechanisms leading to the phenomenon \cite{He2022,Costa2023,Fracassi2024,Coraiola2024,Wang2025,Roy2025a,Sharma2025,Soori2025,Debnath2025,Mondal2025,Shen2025}. Among the major challenges still characterizing the research around the SDE, of particular relevance are the achievement of high efficiencies and the design of platforms for the SDE in the absence of both applied magnetic fields and magnetic ordering.\\
In the present work, we address both aspects from a theoretical perspective. In particular, we focus on proximitized two-dimensional topological insulators (2DTIs) \cite{Bernevig2006a}. Our choice is motivated by several facts, that together make 2DTIs promising for the implementation of the SDE. First, 2DTIs most often emerge in systems with strong spin-orbit coupling \cite{KaneMele2005,Bernevig2006}, making them perfect hosts for states of matter with broken inversion symmetry \cite{Bercioux2015}. Additionally, since they represent ideal candidates for Majorana zero modes \cite{Kitaev2001,Fu2008,Fu2009} and parafermions \cite{Klinovaja2015,Orth2015,Fleckenstein2019}, techniques for their proximization to superconductors have been developed \cite{Hart2014,Wiedenmann2016}. Finally, several experimental realizations are available \cite{Konig2007,Knez2011,Xu2013,Mueller2015,Bismuthene2017,Wu2018,Ghiasi2025}. In particular, we analyze the presence of the SDE in fully proximitized 2DTIs. A schematic of the system under study is shown in Fig. \ref{fig:combined}\textbf{(a)}. The setup is analyzed in two complementary regimes: a wide-sample regime, where the edges do not interact, and a narrow-sample regime, where edge hybridization occurs due to tunneling processes. It is worth to underline that both scenarios have been realized experimentally \cite{Konig2007,strunz}. In the former, we show two ways to realize a SDE with perfect rectification, and in the latter, we demonstrate the presence of a SDE without external fields, realizing an intrinsic SDE. Moreover, we discuss at a microscopic level why the breaking of inversion and time-reversal is not always a sufficient condition to observe SDE in this system.\\\\
Before discussing the phenomenology of the SDE in the proximitized 2DTI setup, we introduce the formalism for calculating the critical current from the energy spectrum in the presence of a current bias, within the non-dissipative phase. This formalism, developed in Ref. \cite{dePicoli2023}, can be summarized as follows.\\ A superconducting current $I_s$ is associated with a finite momentum $2q$ of the Cooper pairs, in particular $I_s = -2q\hbar \rho_s e/m^*$ ($e>0$), where $\rho_s$ is the number of Cooper pairs per unit length, $2q$ corresponds to the momentum of the condensate, and $m^*$ is the effective electron mass. Moreover, we remind that the Ginzburg-Landau equations \cite{Ginzburg1950} provide the additional relation $I_s = (e\hbar\rho_s /m^*)\nabla \varphi$, where $\varphi$ is the superconducting phase.
Correspondingly, the superconducting pairing Hamiltonian for a moving condensate is given by \cite{dePicoli2023}
\begin{gather} 
H_{\Delta}  = \sum_{k}\,\Delta \xi_{k+q\uparrow}^\dagger \xi_{-k+q\downarrow}^\dagger +h.c.,\label{eq:1}
\end{gather}
where $\xi_{k,\uparrow/\downarrow}$ is the Fermi operator with momentum $k$ and spin $\uparrow/\downarrow$, and $\Delta$ is the $k$-independent superconducting gap.
Therefore, the spectrum of the current biased superconducting system explicitly depends on the finite momentum of the Cooper pairs. One can thus define the critical momenta $q_c$ (that correspond to critical currents) as the values of $q$ for which such spectrum becomes gapless. In this framework, a SDE then amounts to the existence of two such momentum values $q_c^+, q_c^-$ with opposite sign and different magnitude. This condition directly encodes the presence of different critical currents injected in opposite spatial directions since, at temperature well below the critical temperature, the Cooper pair density does not depend on the injected current.\\
In this context, non-reciprocity is usually quantified via the rectification coefficient $\eta$, defined as \cite{Strambini2022,Turini2022,Davydova2022,Fracassi2024}
\begin{gather} 
\eta = \frac{|I_c^+|-|I_c^-|}{|I_c^+|+|I_c^-|} \simeq \frac{|q_c^-|-|q_c^+|}{|q_c^+|+|q_c^-|},
\label{eta}
\end{gather}
where in the second equality we made use of the previously introduced critical momenta and the fact that we can neglect the dependence of $\rho_s$ on $I_s$. Below we will use this quantity to evaluate the performance of dissipationless rectification, exploring a large set of parameters. Importantly, we will show both the possibility to achieve unit efficiency and to observe finite SDE ($\eta \neq 0$) without any external magnetic field. 

\section*{Results}\label{sec2}

\subsection*{The microscopic model}
\begin{figure}[h]
  \begin{center}
\raisebox{-0.5\height}{%
  \begin{minipage}{0.42\textwidth}
    \centering
    \includegraphics[width=\linewidth]{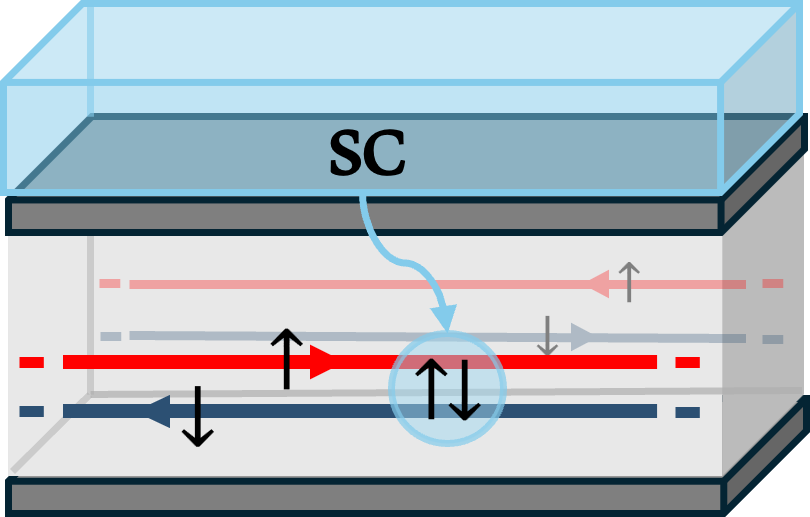}
    \par\smallskip
    \textbf{(a)}
  \end{minipage}
}\hfill
\raisebox{-0.5\height}{%
  \begin{minipage}{0.48\textwidth}
    \centering
    \includegraphics[width=\linewidth]{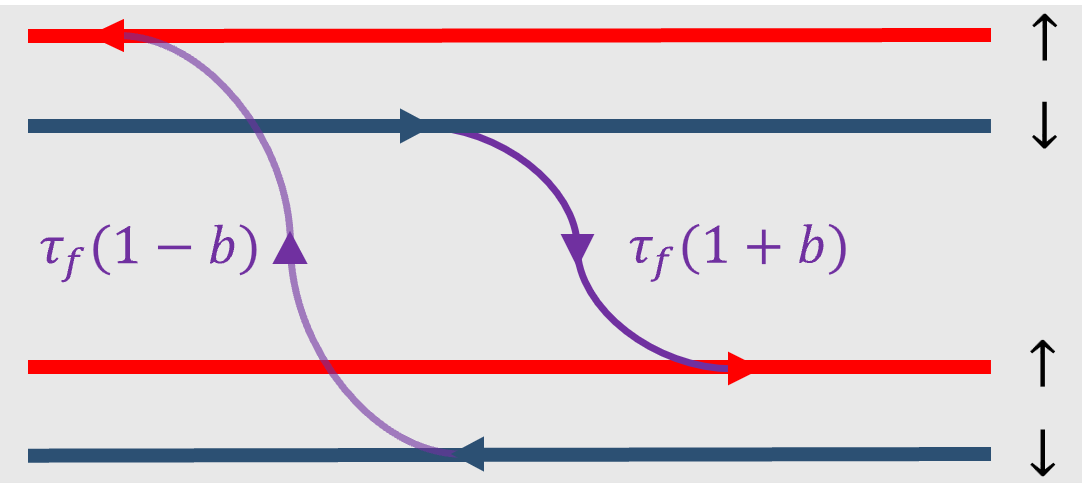}
    \par\smallskip
    \textbf{(b)}
  \end{minipage}
}
\end{center}

  \captionsetup{aboveskip=0pt,belowskip=0pt}
  \caption{{\bf (a) Schematic of the system analyzed}. The superconductor (SC) placed on top of the heterostructure representing the 2DTI induces a superconducting coupling to the two edges. The black arrows represent the spin of the edge states. The colored arrows indicate the direction of motion of spin-polarized edge channels.\\
{\bf (b) Schematics of spin-flip scattering processes and edge reconstruction}. In the sketch we show an example of edge reconstruction with two spin-flip tunneling events (we have sketched the situation for a a fixed $k$ and with $\theta_k=\chi_k$). The tunneling amplitude related to $\tau_f(1+b)c_{k\uparrow}^\dagger d_{k\downarrow} +h.c.$ is favored over the amplitude related to $\tau_f(1-b)c_{k\downarrow}^\dagger d_{k\uparrow} + h.c.$, which implies $b>0$. Here we assumed symmetric spin preserving processes, with $a=0$.}
  \label{fig:combined}
\end{figure}

We now  present the model of a proximitized nanostructure based on the quantum spin Hall phase. The total Hamiltonian is given by
\begin{equation}
H = H_K+H_T+H_B+H_{\Delta},
\label{eq:H}
\end{equation}
where $H_K$ is the kinetic Hamiltonian, $H_T$ models possible tunneling between the opposite edges, $H_B$ describes the coupling with an external magnetic field, and $H_\Delta$ represents the induced superconductivity. Explicitly, we have ($\hbar= 1$)
\begin{equation}
    H_K = \sum_{k, \nu=\pm}\left[(\nu \,v_F k-\mu_u)c_{k\nu}^\dagger c_{k\nu}+(\nu \, v_Fk-\mu_d)d_{k\nu}^\dagger d_{k\nu}\right].
\end{equation}  
Here, $\nu=+/-$ denotes the right/left moving particles, $k$ is the momentum, and $c_{k\nu}$ ($d_{k,\nu}$) a Fermi operator for particles on the upper (lower) edge.
The parameters $\mu_{u/d}$ represent the chemical potentials at the two edges (up/down), whose difference in magnitude can in principle be engineered via side-gate modulation.
The minimal Bernevig-Hughes-Zhang \cite{Bernevig2006} model predicts a $k-$independent spin polarization, so that there is a one to one correspondence between $\nu$ and the spin projection. However, due to spin--orbit coupling, the spin quantization axis is momentum dependent \cite{Liu2008,Konig2008,Ostrovsky2012,Rothe2010,Rothe2014,Virtanen2012,Schmidt2012,Ortiz2016,Ronetti2020}. As a consequence, we can write the following relations for the Fermi operators $c_{k\uparrow/\downarrow}$ and $d_{k\uparrow/\downarrow}$ with well-defined spin
\begin{equation}
    \begin{pmatrix}
    c_{k\uparrow}\\
    c_{k\downarrow}
\end{pmatrix} = \begin{pmatrix}
    \cos \theta_k & -\sin \theta_k\\
    \sin \theta_k & \cos \theta_k
\end{pmatrix}\begin{pmatrix}
    c_{k+}\\
    c_{k-}
\end{pmatrix},
\end{equation}  
\begin{equation}
    \begin{pmatrix}
    d_{k\uparrow}\\
    d_{k\downarrow}
\end{pmatrix} = \begin{pmatrix}
     -\sin \chi_k & \cos \chi_k \\
    \cos \chi_k & \sin \chi_k 
\end{pmatrix}\begin{pmatrix}
    d_{k+}\\
    d_{k-}
\end{pmatrix}.
\end{equation}  
Here $\uparrow/\downarrow$ is the spin-projection along a fixed axis while $+/-$ refers to the direction of propagation. We allow for two distinct mixing angles on the two edges, an effect that can be attributed to a non-uniform Rashba coupling. In principle, in analogy with the behavior of other structures \cite{Affandi2019,Zhang2020,yin2020,Chen2021,Zhi2025}, such non-uniform coupling can be tuned via an external electric field. The angles $\theta_k$ and $\chi_k$ are even functions of $k$ in order to preserve time-reversal symmetry. The case of vanishing angles describes the standard quantum spin Hall effect \cite{KaneMele2005,Bernevig2006}. We emphasize that finite spin mixing ($\theta_k, \chi_k \neq 0$) is an important ingredient for the observation of finite SDE. Indeed, in the setup we analyze, the presence of spin--orbit coupling is mandatory for the manifestation of supercurrent rectification.
It is useful to introduce two effective parameters \(k_{0,u}\) and \(k_{0,d}\), representing the variation scale of the mixing angle for the upper and lower edge, respectively. If this scale is large, as usually assumed \cite{Schmidt2012,Ortiz2016}, the mixing angle can be approximated as $\theta_k \simeq k^2/k_{0,u}^2$ and $\chi_k \simeq k^2/k_{0,d}^2$.
In the spin basis we then have
\begin{gather}
    H_{K} = \sum_k 
\Big(v_F k\cos2\theta_{k}-\mu_u\Big) \, c^\dagger_{k \uparrow} c_{k \uparrow} +\Big(-v_Fk\cos2\theta_{k}-\mu_u\Big) \, c^\dagger_{k \downarrow} c_{k \downarrow} \nonumber\\
+v_Fk \sin2\theta_{k}\, (c^\dagger_{k \uparrow} c_{k \downarrow} + c^\dagger_{k \downarrow} c_{k \uparrow}) \nonumber\\
+\sum_k \Big(-v_Fk\cos2\chi_{k}-\mu_d\Big)\, d^\dagger_{k, \uparrow} d_{k \uparrow} +\Big(v_Fk\cos2\chi_{k}-\mu_d\Big) \, d^\dagger_{k \downarrow} d_{k \downarrow}\nonumber\\
-v_Fk \sin2\chi_{k}\, (d^\dagger_{k \uparrow} d_{k \downarrow} + d^\dagger_{k \downarrow} d_{k \uparrow}).
\end{gather}
If the two edges are close enough, tunneling processes between them can occur. The tunneling Hamiltonian in the spin basis can be written as $H_T = H_p+H_f$, with
\begin{gather}
\label{hp}
    H_p = \sum_k \left[\tau_p(1+a)\,c_{k\uparrow}^\dagger d_{k\uparrow} + \tau_p(1-a)\,c_{k\downarrow}^\dagger d_{k\downarrow} + h.c. \right],
\end{gather}
\begin{gather}
\label{hf}
     H_f=\sum_k\left[\tau_f(1+b)\,c_{k\uparrow}^\dagger d_{k\downarrow} - \tau_f(1-b)\,c_{k\downarrow}^\dagger d_{k\uparrow} + h.c.\right].
\end{gather}
The energy scales $\tau_p$ and $\tau_f$ quantify, respectively, the spin-preserving and spin-flipping tunneling processes between opposite edges. Notice that if $\theta_k, \chi_k =0$, these two terms describe conventional backward and forward tunneling processes \cite{Dolcetto2013,Vigliotti2022}. The two dimensionless coefficients $a$ and $b$ describe edge reconstruction, that is the spatial separation of the two channels on a single edge \cite{recon}. In particular, $a$ ($b$) quantifies time-reversal symmetry breaking that unbalances spin-preserving (spin-flipping) tunneling. In Fig. \ref{fig:combined}\textbf{(b)} we show an example of edge reconstruction for spin-flip tunneling only.
\\
We note that spin-flip tunneling that preserves time-reversal symmetry is justified by the presence of spin--orbit coupling in the bulk \cite{Li2016,Vigliotti2023}.\\
The presence of an externally applied magnetic field is modeled by the Hamiltonian
\begin{equation}
    H_B = \sum_k U_z\left[c_{k\uparrow}^\dagger c_{k\uparrow}-c_{k\downarrow}^\dagger c_{k\downarrow}+d_{k\uparrow}^\dagger d_{k\uparrow}-d_{k\downarrow}^\dagger d_{k\downarrow}\right],
\end{equation}
where $U_z = \frac{1}{2}g \mu_B B$ is the Zeeman energy. Here $g$ is the effective gyromagnetic factor for the edge states, $\mu_B$ the Bohr magneton, and $B$ the magnitude of the magnetic field. 
We consider the Zeeman coupling only, as the large gyromagnetic factor in 2DTIs \cite{Khouri2016} makes this the dominant contribution.\\ The last ingredient of the model is the superconductivity induced at the edges, which is described, in the presence of a current bias as in Eq. \ref{eq:1}, by
\begin{equation}
    H_{\Delta} = \sum_k \Delta\left[c_{k+q\uparrow}^\dagger c_{-k+q\downarrow}^\dagger+d_{k+q\uparrow}^\dagger d_{-k+q\downarrow}^\dagger+h.c.\right],
\end{equation}
where $\Delta$ is assumed to be a real constant and represents the induced gap on the edges.
With the knowledge of the full energy spectrum, we are now in the position of calculating the diode efficiency $\eta$ in the low temperature regime $k_BT\ll \Delta$. By means of exact diagonalization we have numerically evaluated the full energy spectrum as a function of the Cooper pairs momentum $2q$. Then, finding the momenta $q_c^+,q_c^-$ at which the superconducting gap closes, the rectification coefficient $\eta$ is evaluated in different scenarios using Eq. \ref{eta}.\\
It is worth recalling that to observe a finite SDE a necessary (but not sufficient) condition is the simultaneous breaking of time-reversal and inversion symmetry.
Hereafter, we consider two complementary  regimes defined by the width of the heterostructure $w$ (distance between the two edges) with respect to the characteristic penetration length of edge states $\ell$. We first focus on the regime of a large quantum well $w \gg \ell$, where no tunneling between the edges takes place. 
Here, the presence of an external magnetic field explicitly breaks time-reversal symmetry. Since the breaking of spatial inversion symmetry at the level of a single edge has been shown to be insufficient to induce SDE \cite{Dolcini2015,Fracassi2024,Huang2024}, we have introduced an imbalance between the two edges. To do so, we separately consider the effect of an inhomogeneous spin–orbit coupling (thus  \(\theta_k \neq \chi_k\neq 0 \)) and an imbalance in the chemical potential between the two edges ( \(\delta\mu = \mu_d - \mu_u \neq 0\)).\\
Afterwards, we discuss the narrow well geometry $w \lesssim \ell$ where tunneling events play a crucial role. Regarding the breaking of time-reversal symmetry in this regime, we discuss two possibilities. Either the presence of an external magnetic field ($U_z\neq 0$) with bare tunneling events ($a=b=0$) or the presence of edge reconstruction mechanisms ($a,b\neq 0$) without any external fields.
Regarding the breaking of spatial inversion symmetry in the narrow quantum well regime we observe that it is very unlikely to achieve strong differences in mixing angle or chemical potential between the edges. Therefore, in this case we set $\theta_k = \chi_k$, and $\mu_u=\mu_d =\mu$.
The presence of spin-flipping tunneling $\tau_f\neq 0$ guarantees the  breaking of inversion symmetry.

\subsection*{Perfect rectification in large quantum wells}
\begin{figure}[h]
    
  \centering

  \begin{minipage}[c]{0.48\textwidth}
  \centering
  \includegraphics[scale=0.62]{ 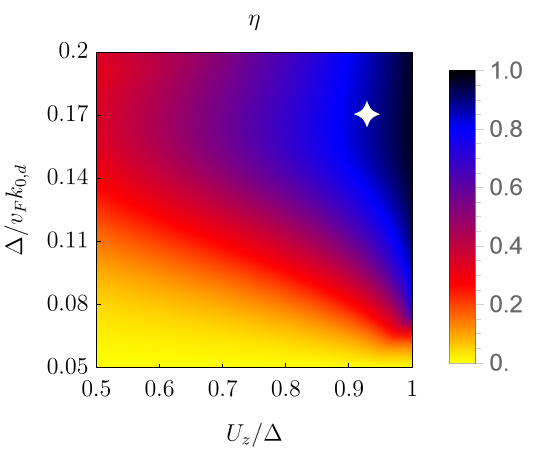}
  \par\smallskip
  \makebox[0pt][c]{\textbf{(a)}}
\end{minipage}\hfill
\begin{minipage}[c]{0.48\textwidth}
  \centering
  \includegraphics[scale=0.32]{ 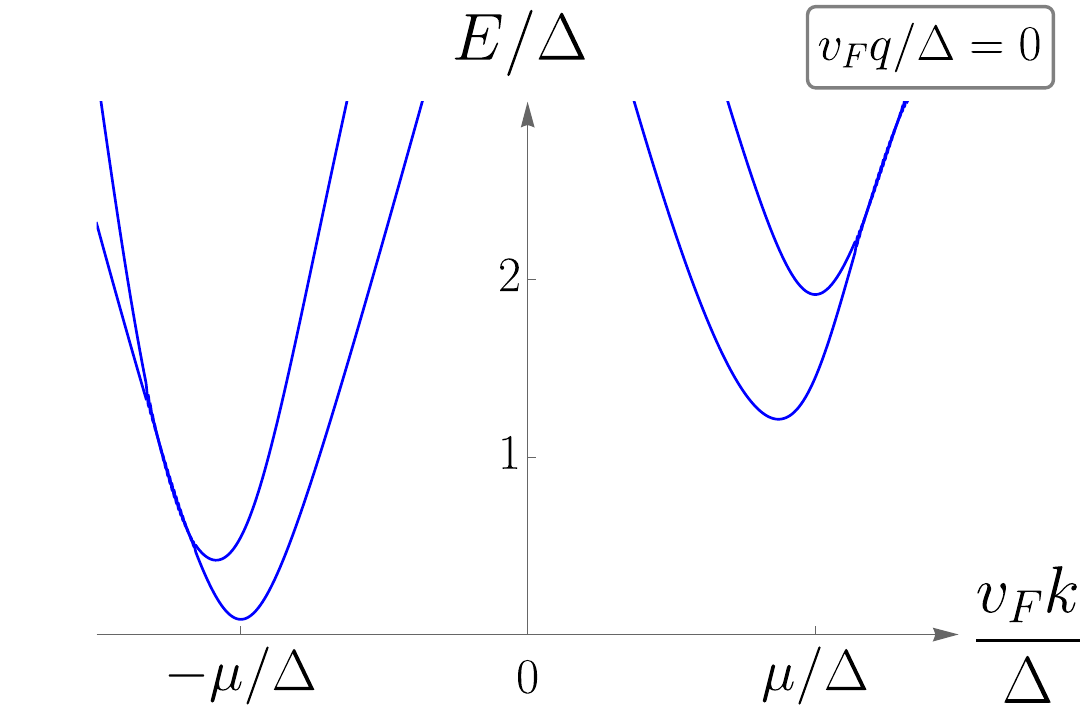}
  \par\smallskip
  \makebox[0pt][c]{\textbf{(b)}}
\end{minipage}

  \vspace{2mm}

  \begin{minipage}[b]{0.48\textwidth}
    \centering
    \includegraphics[scale=0.32]{ 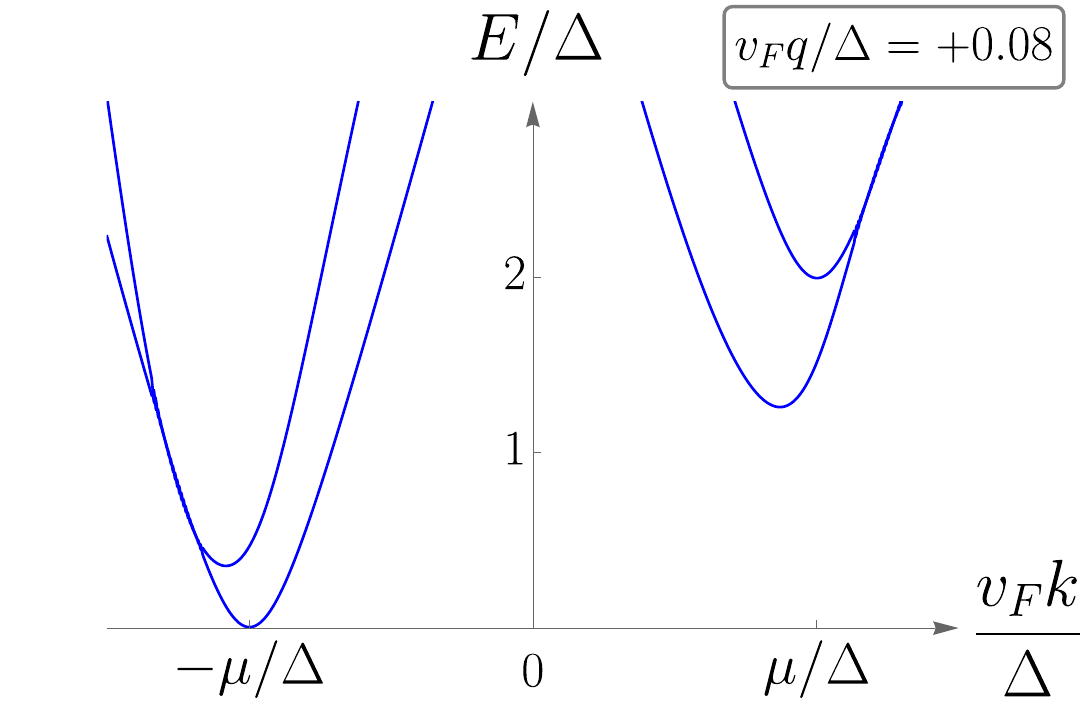}
    \par\smallskip
    \textbf{(c)}\vspace{4pt}
    \label{fig:bands_rqp}
  \end{minipage}\hfill
  \begin{minipage}[b]{0.48\textwidth}
    \centering
    \includegraphics[scale=0.32]{ 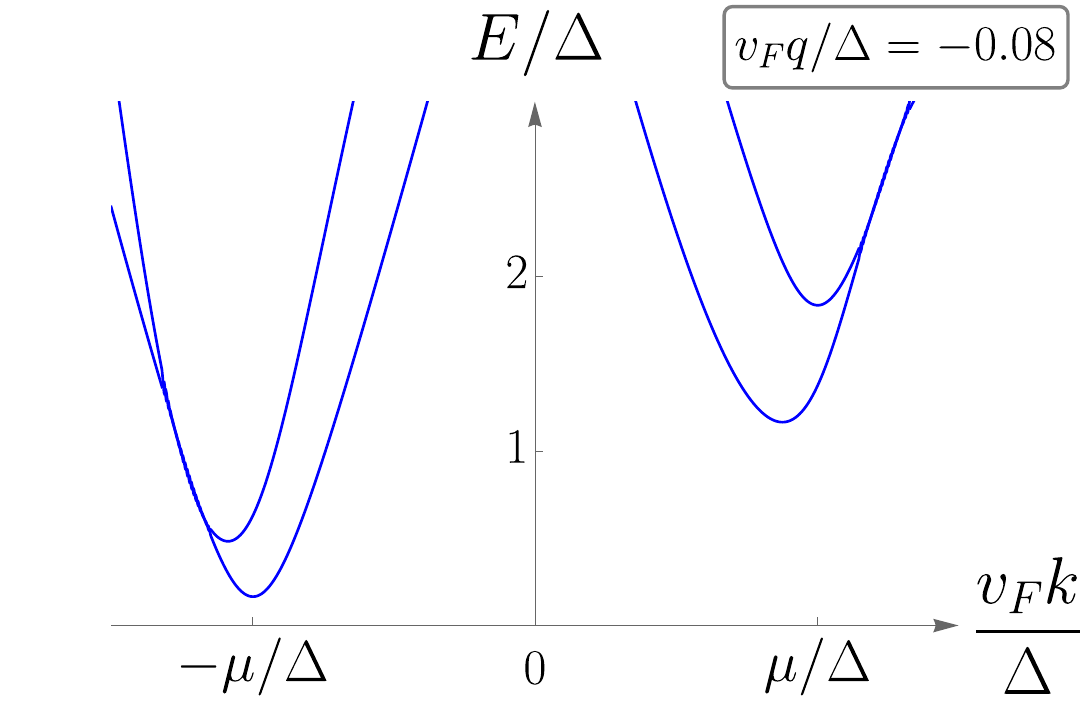}
    \par\smallskip
    \textbf{(d)}\vspace{4pt}
    \label{fig:bands_rqm}
  \end{minipage}
    \caption{{\bf SDE performance with inhomogeneous spin--orbit coupling.} The panel \textbf{(a)} shows the rectification coefficient as a function of magnetic field $U_z/\Delta$ and Rashba inhomogeneity $\Delta/v_Fk_{0,d}$. A near unit efficiency can be achieved for $U_z \lesssim \Delta$. Here, $\Delta/v_Fk_{0,u} = 0.05$ and $\mu/\Delta = 6$. The other three panels show the bands with different current bias at the parameters corresponding to the white star ($U_z/\Delta=0.93$, $\Delta/v_Fk_{0,d}=0.17$) in panel \textbf{(a)}. Panel \textbf{(b)} shows the unbiased bands illustrating how symmetry breaking affects the bands. Since the superconducting gap corresponding to $-\mu/\Delta$ is very close to zero, it can be closed by a small $q>0$. This is shown in panel \textbf{(c)}. Injecting the same amount of current in the opposite direction (panel \textbf{(d)}) does not close any gap, and this lead to a rectification of the supercurrent with $\eta \simeq 1$.}
    \label{fig4}
\end{figure}

\begin{figure}[h]

 \centering

  \begin{minipage}[c]{0.48\textwidth}
  \centering
  \includegraphics[scale=0.62]{ 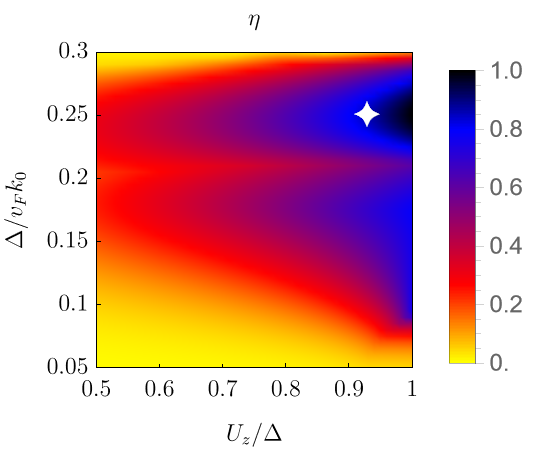}
  \par\smallskip
  \makebox[0pt][c]{\textbf{(a)}}
\end{minipage}\hfill
\begin{minipage}[c]{0.48\textwidth}
  \centering
  \includegraphics[scale=0.32]{ 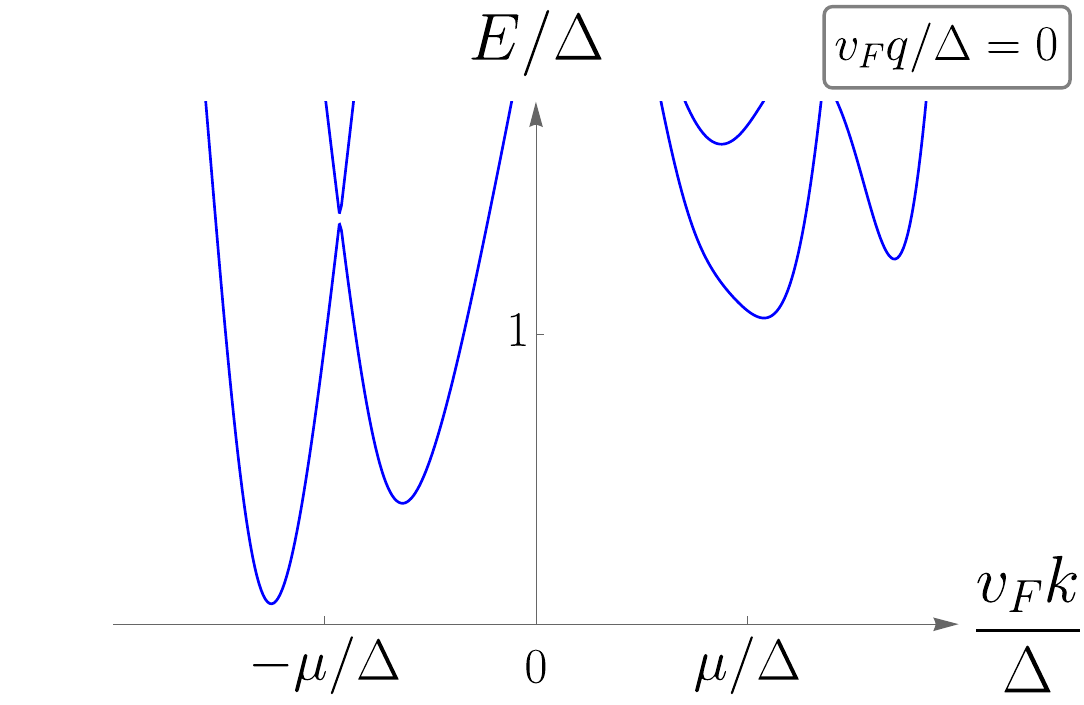}
  \par\smallskip
  \makebox[0pt][c]{\textbf{(b)}}
\end{minipage}

  \vspace{2mm}

  \begin{minipage}[b]{0.48\textwidth}
    \centering
    \includegraphics[scale=0.32]{ 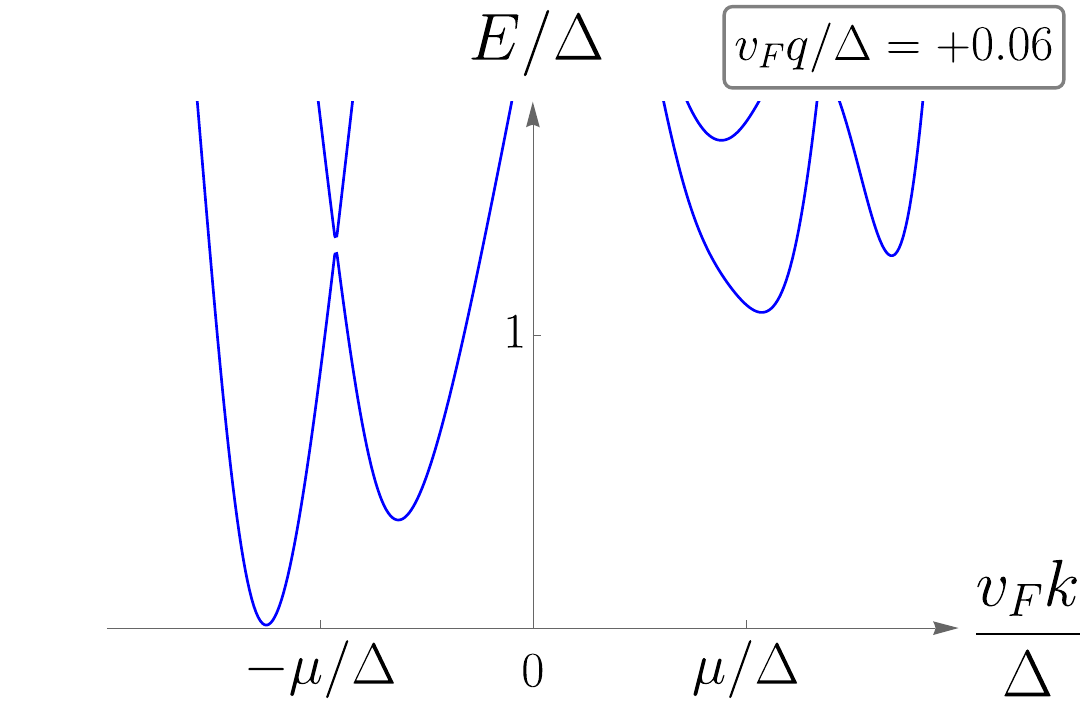}
    \par\smallskip
    \textbf{(c)}\vspace{4pt}
    \label{fig:bands_rqp}
  \end{minipage}\hfill
  \begin{minipage}[b]{0.48\textwidth}
    \centering
    \includegraphics[scale=0.32]{ 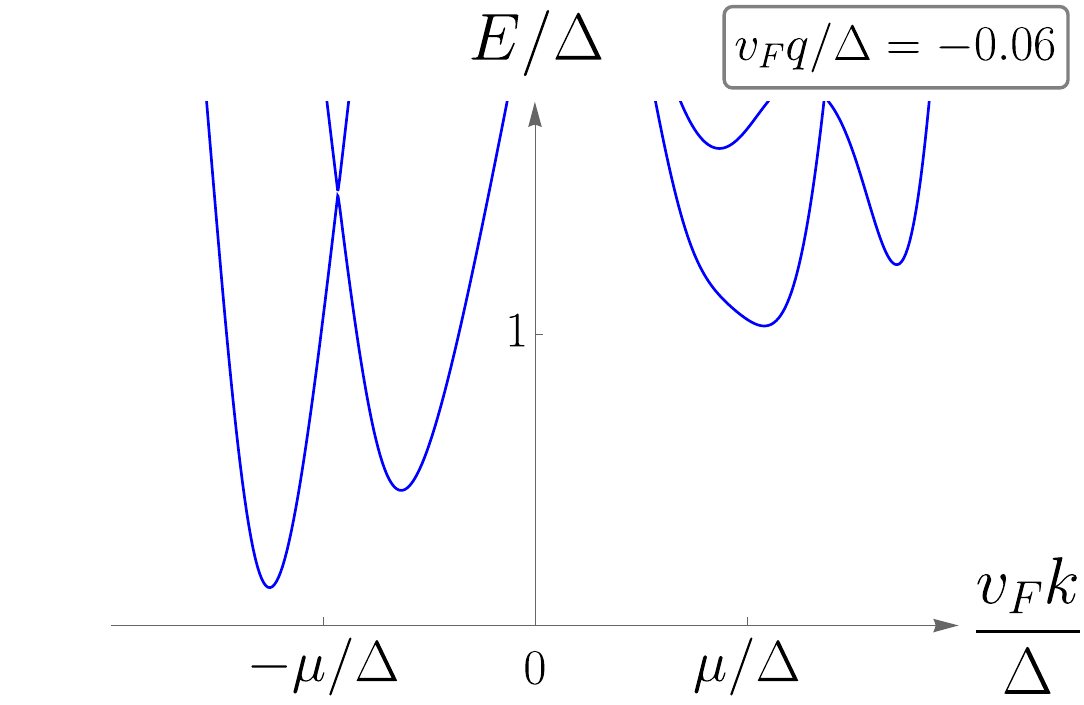}
    \par\smallskip
    \textbf{(d)}\vspace{4pt}
    \label{fig:bands_rqm}
  \end{minipage}
    \centering
    \caption{{\bf SDE performance with inhomogeneous chemical potentials.}
    Panel \textbf{(a)} shows the rectification coefficient as a function of magnetic field $U_z/\Delta$ and spin--orbit coupling $\Delta/v_Fk_{0}$. A near unit efficiency can be achieved for $U_z \lesssim \Delta$ and $\Delta/v_Fk_{0} \simeq 0.25$. Here, $\delta\mu/\Delta=2, \mu/\Delta = 4$. The remaining three panels display the band structures under different current bias for the parameters indicated by the white star (\(U_z/\Delta=0.93\), \(\Delta/v_F k_{0}=0.25\)) in panel \textbf{(a)}. Panel \textbf{(b)} shows the unbiased bands and highlights the effect of symmetry breaking. Because the superconducting gap near \(-\mu/\Delta\) is almost zero, a small positive momentum \(q>0\) is sufficient to close it, as shown in panel \textbf{(c)}. Injecting the same magnitude of current in the opposite direction (panel \textbf{(d)}) does not close any gap, leading to near‑perfect rectification of the supercurrent, \(\eta \simeq 1\).
}
    \label{panel2}
\end{figure}

We now discuss the large well regime $w \gg \ell$, where tunneling processes between the two edges can be safely neglected. In terms of Hamiltonian parameters this means $\tau_p=\tau_f=0$.
First we consider the case of non-homogeneous spin-orbit coupling, i.e., $\theta_k\neq \chi_k$ (or equivalently $1/k_{0,u}\neq 1/k_{0,d}$), but equal edge chemical potentials $\mu_u=\mu_d=\mu$.
Hereafter, the induced gap $\Delta$ is adopted as the reference energy scale. Moreover, $1/v_Fk_{0,u}$ is kept constant at $\Delta/v_Fk_{0,u}=0.05$, so that $1/v_Fk_{0,d}$ directly quantifies the strength of the spin--orbit inhomogeneity.
The results are reported in Fig. \ref{fig4}, where the rectification coefficient $\eta$ is shown as a function of Zeeman energy $U_z$ and Rashba inhomogeneity $1/v_Fk_{0,d}$ at a fixed chemical potential $\mu=6\Delta$ (see also the supplementary information for more details).
In Fig. \ref{fig4}\textbf{(a)} one can observe that the rectification grows monotonically with magnetic field reaching values close to unity for $U_z \lesssim \Delta$, and thus realizing a perfect SDE (unit efficiency).
Looking at the energy bands without current bias ($q=0$) shown in Fig. \ref{fig4}\textbf{(b)}, this perfect rectification can be attributed to a near-zero value of one of the two superconducting gaps, which is closed by a small current (exactly zero in the limiting case of critical Zeeman energy $U_{z,c}=\Delta$) but only for one of the two injection directions (as shown in Fig. \ref{fig4}\textbf{(c)}). The other superconducting gap is much larger and does not close with the same amount of current flowing in the opposite direction (see the energy spectrum of Fig. \ref{fig4}(\textbf{d})), and this leads to $\eta \simeq 1$.
\\
Fig. \ref{panel2} shows the results for a chemical potential unbalance $\delta \mu=\mu_d-\mu_u$, with an homogeneous spin-orbit coupling $1/k_{0,u}=1/k_{0,d}=1/k_0$ (see also supplementary material for additional details). In panel ($\mathbf{a}$), $\eta$ is shown as a function of $1/v_Fk_0$ and $U_z$ at fixed $\delta\mu=2\Delta$ and $\mu= (\mu_u+\mu_d)/2=4\Delta$. Interestingly, also in this physical situation $\eta \simeq 1$ can be achieved. The leading mechanism is analog to the one discussed above for the inhomogeneous spin--orbit coupling case, as shown by the energy bands in Fig. \ref{panel2}. 
\subsection*{Intrinsic SDE in narrow quantum wells}

\begin{figure}[h]

 \centering

  \begin{minipage}[c]{0.48\textwidth}
  \centering
  \includegraphics[scale=0.62]{ 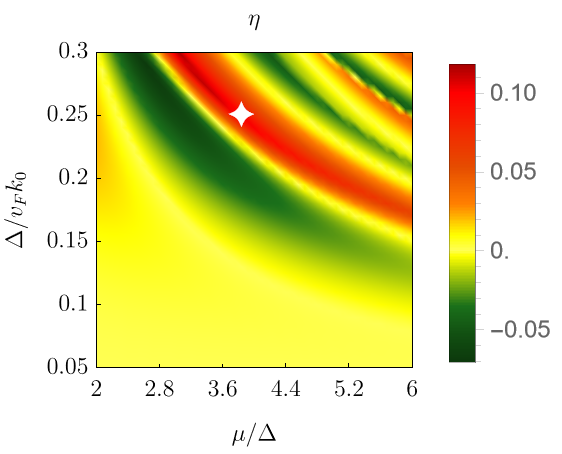}
  \par\smallskip
  \makebox[0pt][c]{\textbf{(a)}}
\end{minipage}\hfill
\begin{minipage}[c]{0.48\textwidth}
  \centering
  \includegraphics[scale=0.32]{ 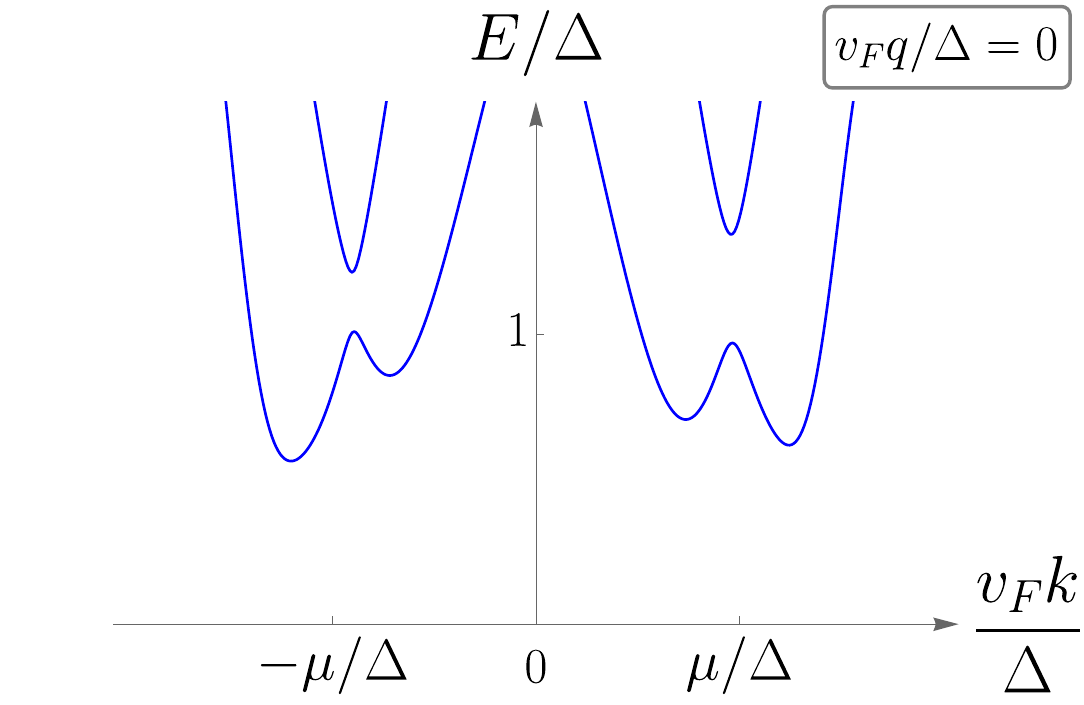}
  \par\smallskip
  \makebox[0pt][c]{\textbf{(b)}}
\end{minipage}

  \vspace{2mm}

  \begin{minipage}[b]{0.48\textwidth}
    \centering
    \includegraphics[scale=0.32]{ 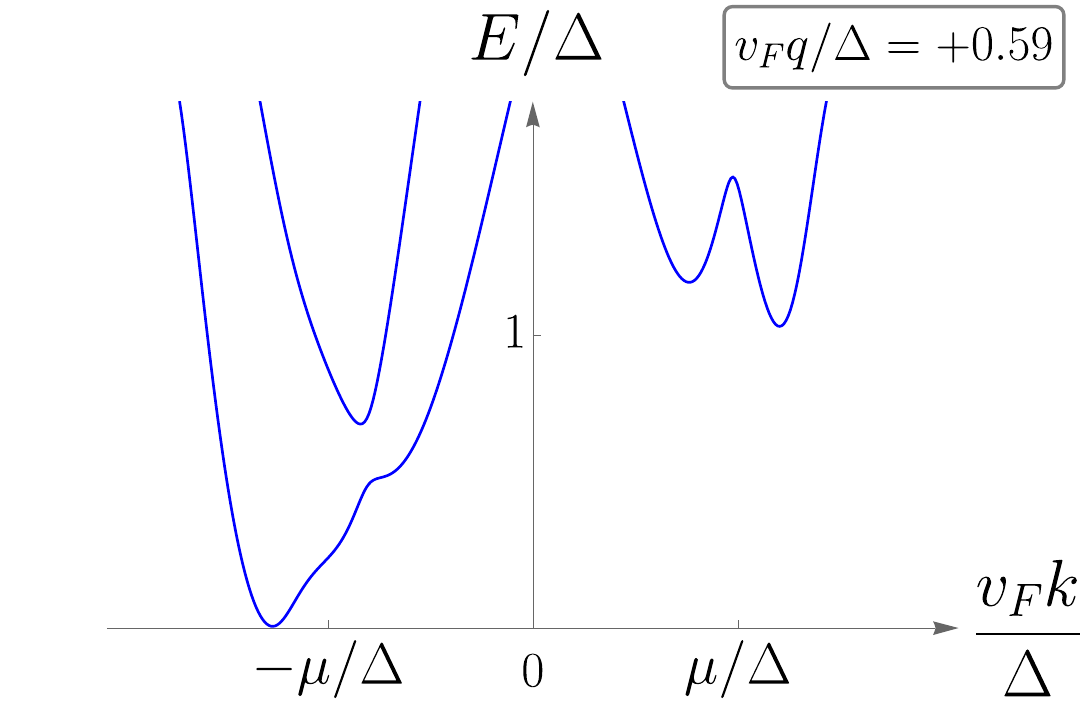}
    \par\smallskip
    \textbf{(c)}\vspace{4pt}
    \label{fig:bands_rqp}
  \end{minipage}\hfill
  \begin{minipage}[b]{0.48\textwidth}
    \centering
    \includegraphics[scale=0.32]{ 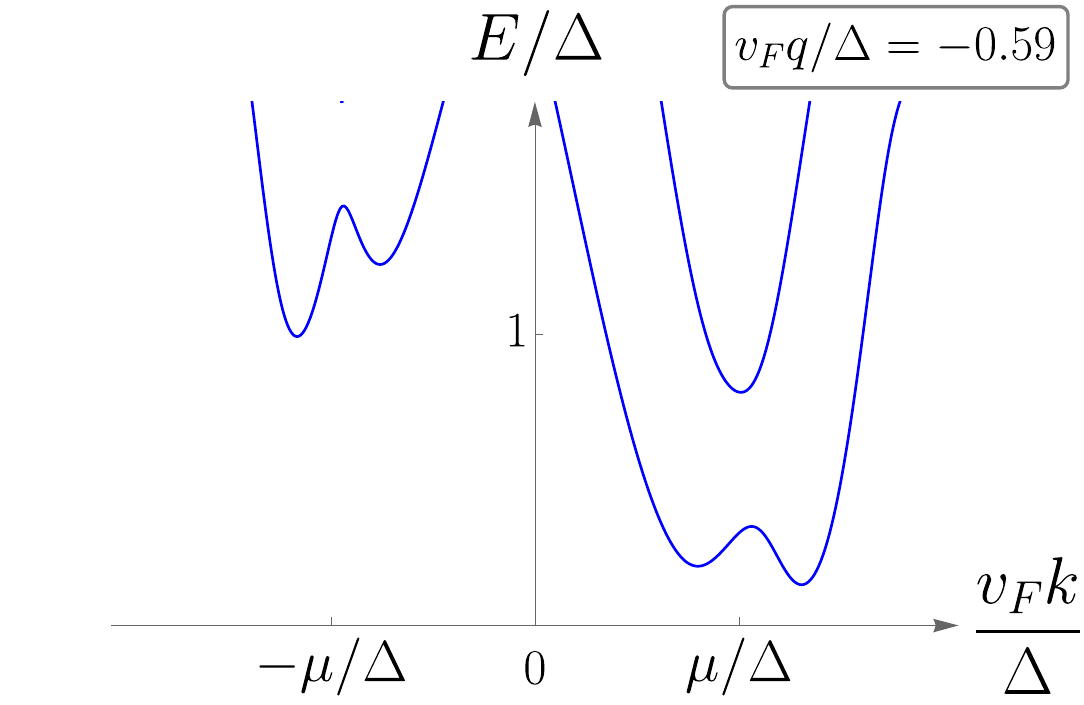}
    \par\smallskip
    \textbf{(d)}\vspace{4pt}
    \label{fig:bands_rqm}
  \end{minipage}
    \caption{{\bf SDE performance in the narrow well regime without edge reconstruction ($a=b=0$).} Panel \textbf{(a)} shows the rectification coefficient as a function of spin--orbit strength $\Delta/v_Fk_0$ and chemical potential $\mu/\Delta$. Here, $U_z/\Delta = 0.85,  \tau_f/\Delta = 0.6, \tau_p/\Delta=0.7$. The remaining three panels illustrate the band structures under different current biases for the parameters marked by the white star (\(\mu/\Delta = 3.85\), \(\Delta/v_F k_{0} = 0.25\)). Panel \textbf{(b)} displays the unbiased bands, highlighting the impact of symmetry breaking (the difference between the values of the two gaps at $q=0$ in this case is smaller but is not zero). A small positive momentum \(q>0\) suffices to fully close the superconducting gap near \(-\mu/\Delta\), as shown in panel \textbf{(c)}. Applying the same current in the opposite direction (panel \textbf{(d)}) leaves the gap open, resulting in rectification of the supercurrent.}
    \label{panel3}
\end{figure}

\begin{figure}[h]
 \centering
 
  \begin{minipage}[c]{0.48\textwidth}
  \centering
  \includegraphics[scale=0.62]{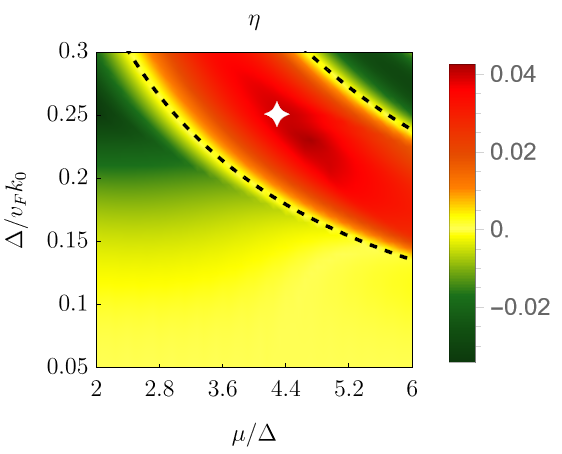}
  \par\smallskip
  \makebox[0pt][c]{\textbf{(a)}}
\end{minipage}\hfill
\begin{minipage}[c]{0.48\textwidth}
  \centering
  \includegraphics[scale=0.32]{ 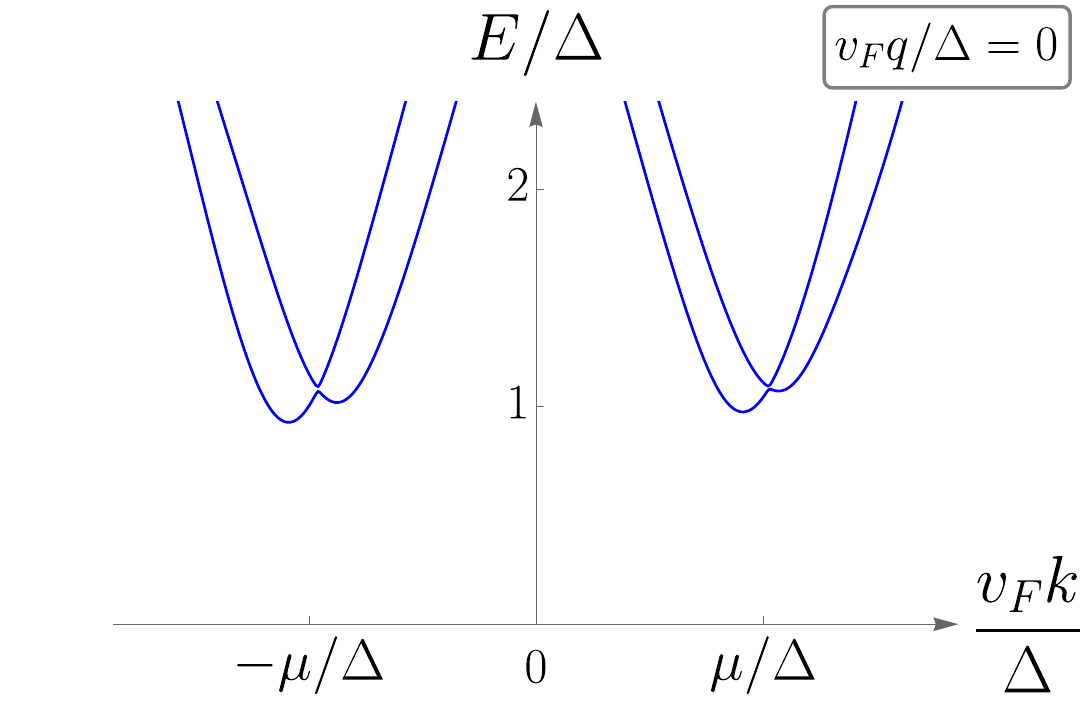}
  \par\smallskip
  \makebox[0pt][c]{\textbf{(b)}}
\end{minipage}

  \vspace{2mm}

  \begin{minipage}[b]{0.48\textwidth}
    \centering
    \includegraphics[scale=0.32]{ 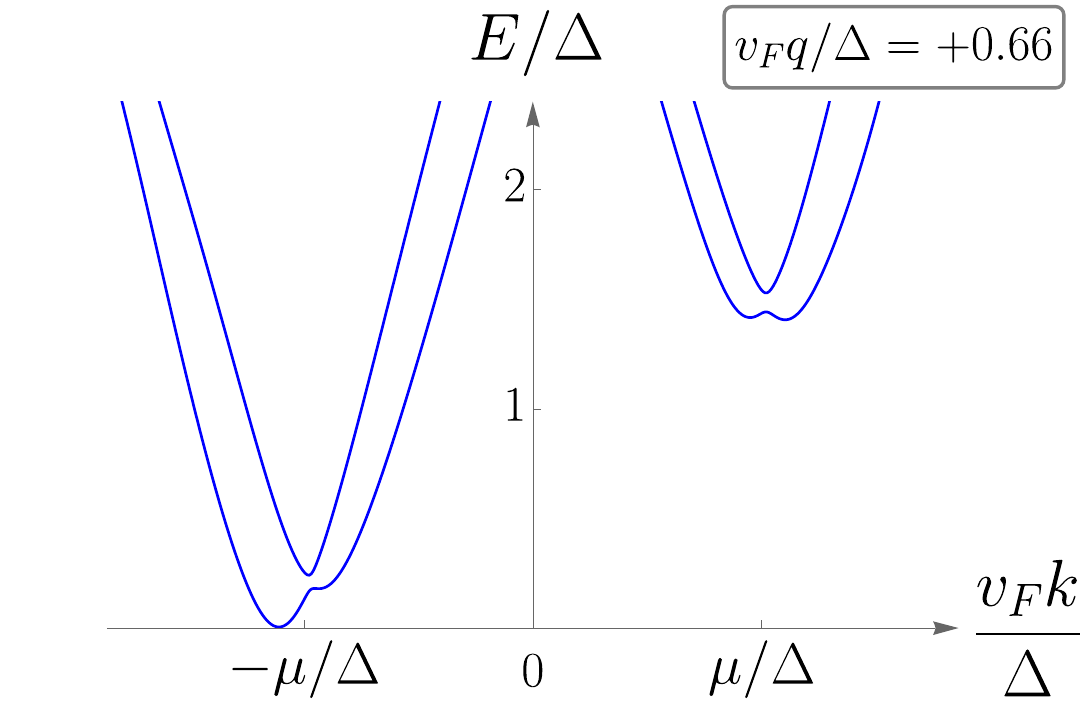}
    \par\smallskip
    \textbf{(c)}\vspace{4pt}
    \label{fig:bands_rqp}
  \end{minipage}\hfill
  \begin{minipage}[b]{0.48\textwidth}
    \centering
    \includegraphics[scale=0.32]{ 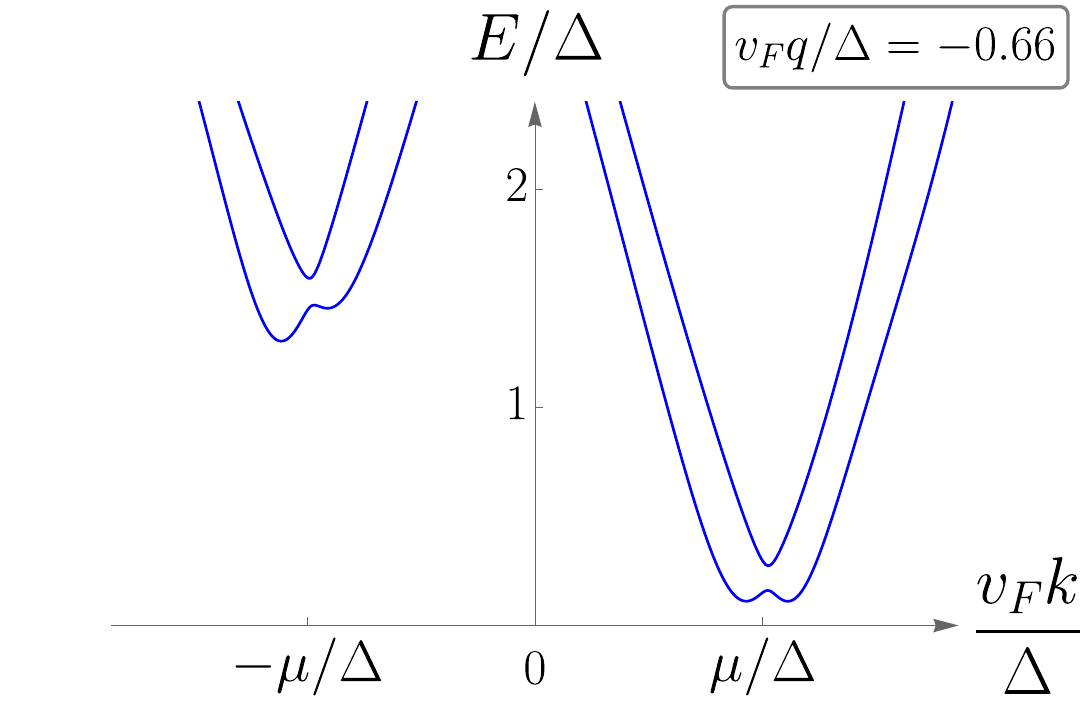}
    \par\smallskip
    \textbf{(d)}\vspace{4pt}
    \label{fig:bands_rqm}
  \end{minipage}
    \caption{{\bf Intrinsic SDE in the narrow well regime with edge reconstruction.}
    Panel \textbf{(a)} presents the rectification coefficient as a function of spin--orbit strength $\Delta/v_Fk_0$, and chemical potential $\mu/\Delta$. Here, we have set $b=0.2, a=0$ $\tau_f/\Delta = 0.4, \tau_p/\Delta = 0.7$. The remaining three panels show the band structures under different values of current bias for the parameters indicated by the white star (\(\mu/\Delta = 4.30\), \(\Delta/v_F k_{0} = 0.25\)) in panel \textbf{(a)}. Panel \textbf{(b)} depicts the unbiased bands, emphasizing the role of symmetry breaking. A positive momentum \(q>0\) is sufficient to close the superconducting gap near \(-\mu/\Delta\), as illustrated in panel \textbf{(c)}. Driving the same current in the opposite direction (panel \textbf{(d)}) leaves the gap open, thereby producing rectification of the supercurrent. In panel \textbf{(a)} we observe zero rectification at specific values of the couple $(\mu,1/v_Fk_0)$ despite the breaking of the symmetries. The dashed lines represent our theoretical explanation of why along these lines the rectification is zero despite the breaking of the two necessary symmetries (see supplementary information).}
    \label{panel4}
\end{figure}

We now discuss the case when the edges are spatially separated by a distance smaller than their penetration length, i.e. $w\lesssim \ell$. In particular, we will concentrate on how tunneling processes can affect SDE. It should first be emphasized that the presence of weak tunneling does not qualitatively alter the physics presented in the previous Section, see the supplementary material for further details. However, the narrow well regime also offers additional possibilities to observe finite SDE. We now illustrate the effect of a magnetic field and tunneling without edge reconstruction ($a=b=0$ in Eqs. \ref{hp}-\ref{hf}).
 Fig. \ref{panel3}\textbf{(a)} shows the dependence of $\eta$ on $1/v_Fk_{0,u}=1/v_Fk_{0,d}\equiv1/v_Fk_0$ and $\mu$ at fixed $U_z$, $\tau_f$, and $\tau_p$, showing a non-monotonic modulation of the rectification coefficient (see also supplementary material for other parameter configurations). 
\\
The band structure that leads to SDE is shown in Fig. \ref{panel3}\textbf{(b)}. It is worth to underline that for $\tau_p=0$, no SDE occurs (not shown), even though all the necessary symmetry breakings are present.\\
We now look for an intrinsic SDE, thus without any external magnetic field, i.e., $U_z=0$. As stated above, we rely on edge reconstruction mechanism to break time-reversal symmetry, and without loss of generality we consider $0< a,b \leq 1$ in the following.
In Fig. \ref{panel4} we show rectification for the simultaneous presence of reconstructed spin-flip tunneling ($b\neq 0$) and non-reconstructed spin-preserving tunneling ($a=0$). 
In this scenario, a (homogeneous) spin–-orbit coupling strength of at least $\Delta/v_Fk_0 \simeq 0.1$ is required in order to observe a finite rectification. 
Although the transition is not sharp, this provides clear evidence that a substantial spin--orbit interaction is essential for the emergence of the intrinsic SDE. Fig. \ref{panel4}$\mathbf{(a)}$ illustrates $\eta$ as a function of $1/v_Fk_0$ and $\mu$ at fixed $b=0.2$. Interestingly, a region of zero rectification is evident for specific values of the couple $(\mu,1/v_Fk_0)$ despite symmetry breaking. This behavior is independent of $b$ (the independence on $b$ and our theoretical explanation are reported in the supplementary information).
In panel \textbf{(b)} we show the band structure that leads to SDE. 
In passing, we note that in this regime the SDE emerges even in the absence of spin-preserving tunneling (see the supplementary information), but in order to present results in a physically relevant region of the parameter space, we have kept $\tau_p \neq 0$ in Fig.~\ref{panel4}.
The situation with $a\neq0,b=0$ (and a combination of the two) also manifests SDE, as shown in the supplementary information.

\section*{Discussion}\label{sec12}
We have shown the possibility to achieve finite SDE in proximitized quantum spin Hall heterostructures in a wide range of parameters. In particular, we focused on two distinct regimes: large quantum wells, where no tunneling between opposite edges occurs, and narrow ones where tunneling processes play a crucial role. In the former case, in the presence of an external magnetic field, we have shown the possibility to achieve rectification coefficients close to unity. In the latter, instead, we observed finite SDE even without any external field, due to spin-flip tunneling events with edge reconstruction that intrinsically break time-reversal symmetry.\\
As expected, the breaking of time-reversal and inversion symmetries is not always sufficient to generate SDE. In all cases analyzed, the breaking of the symmetries must be accompanied by two additional conditions: (i) an asymmetry of the superconducting gaps in $k$-space in the absence of a current bias, and (ii) a $k$-dependence in the spin pattern of the excitations above the superconducting gap. A deeper discussion about these two conditions is reported in the supplementary material.\\
It is worth to mention another consequence of our results. Despite the fact that our model undergoes a topological phase transition associated with the appearance of Majorana zero energy modes in the finite structure setup \cite{trenches}, no signatures of such a topological quantum phase transitions are present in the behavior of the rectification coefficient. This fact represents a word of caution in terms of distinguishing the platforms and perspectives of the superconducting Josephson diode effect and the SDE in translational invariant systems.\\
To conclude, we briefly comment on the experimental relevance of our results. We argue that
CdTe/HgTe quantum wells within the helical regime can be a good playground to observe these effects. Indeed, the critical current measured in \cite{Hart2014} leads to the estimate $\Delta \simeq \sqrt{\frac{\pi\hbar m^* v_F^2}{4e}\SI{10}{\nano\ampere}} = \SI{140}{\micro\electronvolt}$ , and since the effective magnetic moment of the electrons in this system is $g \simeq 40$ \cite{Khouri2016}, a typical value for the magnetic field is given by $\SI{63}{\milli\tesla} = \frac{\Delta}{g\mu_B} \leq B \leq \frac{2\Delta}{g\mu_B} = \SI{126}{\milli\tesla}$. Another condition to be satisfied is that the $k$ range $\Delta k_{\text{low-E}}$ involved in the low-energy physics is contained within the range $\Delta k_{\text{HgTe}}$ where linear bands are present before connecting to the bulk bands. The $k$ values involved in the low-energy physics range from $-\mu$ to $\mu$, which is an estimate of the $k$ value at which the superconducting gap closes. The maximum value of $\mu$ analyzed is $6\Delta = \SI{840}{\micro\electronvolt}$, leading to $\Delta k_{\text{low-E}} \simeq \SI{26}{\per\micro\metre}$. Since one can estimate $\Delta k_{\text{HgTe}} \simeq \SI{150}{\per\micro\metre}$  \cite{Bernevig2006}, the condition $\Delta k_{\text{low-E}} < \Delta k_{\text{HgTe}}$ is satisfied in all cases of interest. Finally, the reported values of $\hbar v_F k_0/\Delta$ ensure that in the $k$ range where the analyzed physics takes place, at most one complete oscillation of the spin mixing appears.
All these considerations allow us to conclude that the CdTe/HgTe heterostructure is a promising platform for observing a perfect SDE in the wide structure regime, and that it is also possible to observe the intrinsic SDE in the regime of tunneling-hybridized edges, which would represent a strong indicator of edge reconstruction physics in the quantum spin Hall regime.

\section*{Acknowledgements}

The authors acknowledge support from the Project PRIN2022 2022-PH852L(PE3) TopoFlags—“Non reciprocal supercurrent and
topological transition in hybrid Nb-InSb nanoflags” funded by the
European community’s Next Generation EU within the program
“PNRR Missione 4—Componente 2—Investimento 1.1 Fondo per il
Programma Nazionale di Ricerca e Progetti di Rilevante Interesse
Nazionale (PRIN).” S.H. acknowledges partial support by PNRR
MUR, Project No. PE0000023-NQSTI.

\printbibliography

@article{Hu2007,
  title = {Proposed Design of a Josephson Diode},
  author = {Hu, J. and Wu, C. and Dai, X.},
  journal = {Phys. Rev. Lett.},
  volume = {99},
  pages = {067004},
  year = {2007}
}

@article{Wakatsuki2017,
  title = {Nonreciprocal charge transport in noncentrosymmetric superconductors},
  author = {Wakatsuki, R. and Saito, Y. and Hoshino, S. and Itahashi, Y. M. and Ideue, T. and Ezawa, M. and Iwasa, Y. and Nagaosa, N.},
  journal = {Sci. Adv.},
  volume = {3},
  pages = {e1602390},
  year = {2017}
}

@article{Hoshino2018,
  title = {Nonreciprocal charge transport in two-dimensional noncentrosymmetric superconductors},
  author = {Hoshino, S. and Wakatsuki, R. and Hamamoto, K. and Nagaosa, N.},
  journal = {Phys. Rev. B},
  volume = {98},
  pages = {054510},
  year = {2018}
}

@article{Yasuda2019,
  title = {Nonreciprocal charge transport at topological insulator/superconductor interface},
  author = {Yasuda, K. and Yasuda, H. and Liang, T. and Yoshimi, R. and Tsukazaki, A. and Takahashi, K. S. and Nagaosa, N. and Kawasaki, M. and Tokura, Y.},
  journal = {Nat. Commun.},
  volume = {10},
  pages = {2734},
  year = {2019}
}

@article{Ando2020,
  title = {Observation of superconducting diode effect},
  author = {Ando, F. and Miyasaka, Y. and Li, T. and Ishizuka, J. and Arakawa, T. and Shiota, Y. and Moriyama, T. and Yanase, Y. and Ono, T.},
  journal = {Nature},
  volume = {584},
  pages = {373},
  year = {2020}
}

@article{Misaki2021,
  title = {Theory of the nonreciprocal Josephson effect},
  author = {Misaki, K. and Nagaosa, N.},
  journal = {Phys. Rev. B},
  volume = {103},
  pages = {245302},
  year = {2021}
}

@article{Tanaka2022,
  title = {Theory of giant diode effect in d-wave superconductor junctions on the surface of a topological insulator},
  author = {Tanaka, Y. and Lu, B. and Nagaosa, N.},
  journal = {Phys. Rev. B},
  volume = {106},
  pages = {214524},
  year = {2022}
}

@article{Daido2022,
  title = {Intrinsic Superconducting Diode Effect},
  author = {Daido, A. and Ikeda, Y. and Yanase, Y.},
  journal = {Phys. Rev. Lett.},
  volume = {128},
  pages = {037001},
  year = {2022}
}

@article{Lu2023,
  title = {Tunable Josephson Diode Effect on the Surface of Topological Insulators},
  author = {Lu, B. and Ikegaya, S. and Burset, P. and Tanaka, Y. and Nagaosa, N.},
  journal = {Phys. Rev. Lett.},
  volume = {131},
  pages = {096001},
  year = {2023}
}

@article{Reinhardt2024,
  title = {Ubiquitous Superconducting Diode Effect in Superconductor Thin Films},
  author = {Reinhardt, S. and Ascherl, T. and Costa, A. and Berger, J. and Gronin, S. and Gardner, G. C. and Lindemann, T. and Manfra, M. J. and Fabian, J. and Kochan, D. and Strunk, C. and Paradiso, N.},
  journal = {Nat. Commun.},
  volume = {15},
  pages = {4413},
  year = {2024}
}

@article{Souto2024,
  title = {Tuning the Josephson Diode Response with an AC Current},
  author = {Souto, R. S. and Leijnse, M. and Schrade, C. and Valentini, M. and Katsaros, G. and Danon, J.},
  journal = {Phys. Rev. Res.},
  volume = {6},
  pages = {L022002},
  year = {2024}
}

@article{Cayao2024,
  title = {Enhancing the Josephson Diode Effect with Majorana Bound States},
  author = {Cayao, J. and Nagaosa, N. and Tanaka, Y.},
  journal = {Phys. Rev. B},
  volume = {109},
  pages = {L081405},
  year = {2024}
}

@article{KaneMele2005,
  title = {Quantum Spin Hall Effect in Graphene},
  author = {Kane, C. L. and Mele, E. J.},
  journal = {Phys. Rev. Lett.},
  volume = {95},
  pages = {226801},
  year = {2005}
}

@article{Vigliotti2022,
  title = {Anomalous flux periodicity in proximitised quantum spin Hall constrictions},
  author = {Vigliotti, L. and Calzona, A. and Trauzettel, B. and Sassetti, M. and Ziani, N. T.},
  journal = {New J. Phys.},
  volume = {24},
  pages = {053017},
  year = {2022}
}

@article{Fracassi2024,
  title = {Anomalous supercurrent and diode effect in locally perturbed topological Josephson junctions},
  author = {Fracassi, S. and Traverso, S. and Ziani, N. T. and Carrega, M. and Heun, S. and Sassetti, M.},
  journal = {Appl. Phys. Lett.},
  volume = {124},
  pages = {242601},
  year = {2024}
}

@article{Huang2024,
  title = {Superconducting diode effect in two-dimensional topological insulator edges and Josephson junctions},
  author = {Huang, H. and de Picoli, T. and Väyrynen, J. I.},
  journal = {Appl. Phys. Lett.},
  volume = {125},
  pages = {032602},
  year = {2024}
}

@article{Sharma2025,
  title = {Tunable Josephson diode effect in singlet superconductor-altermagnet-triplet superconductor junctions},
  author = {Sharma, L. and Thakurathi, M.},
  journal = {Phys. Rev. B},
  volume = {112},
  pages = {104506},
  year = {2025}
}

@article{Soori2025,
  title = {Josephson diode effect in one-dimensional quantum wires connected to superconductors with mixed singlet-triplet pairing},
  author = {Soori, A.},
  journal = {J. Phys.: Condens. Matter},
  volume = {37},
  pages = {10LT02},
  year = {2025}
}

@article{Debnath2025,
  title = {Field-free Josephson diode effect in interacting chiral quantum dot junctions},
  author = {Debnath, D. and Dutta, P.},
  journal = {J. Phys.: Condens. Matter},
  volume = {37},
  pages = {175301},
  year = {2025}
}

@article{Mondal2025,
  title = {Josephson diode effect with Andreev and Majorana bound states},
  author = {Mondal, S. and Fu, P.-H. and Cayao, J.},
  journal = {Phys. Rev. B},
  volume = {112},
  pages = {085401},
  year = {2025}
}

@article{Shen2025,
  title = {Josephson diodes induced by loop current states},
  author = {Shen, Q. K. and Zhang, Y.},
  journal = {Phys. Rev. B},
  volume = {111},
  pages = {174515},
  year = {2025}
}

@article{Roy2025a,
  title = {Floquet-engineered diode performance of a topological Josephson junction composed of two Kitaev chains coupled via a quantum dot},
  author = {Roy, K. and Paul, G. and Debnath, D. and Bhattacharyya, K. and Basu, S.},
  journal = {Phys. Rev. B},
  volume = {112},
  pages = {125407},
  year = {2025}
}

@article{Wang2025,
  title = {Theoretical study of superconducting diode effect in planar T$_d$--MoTe$_2$ Josephson junctions},
  author = {Wang, G.-Q. and Miao, J.-J. and Chen, W.-Q.},
  journal = {Phys. Rev. B},
  volume = {112},
  pages = {014508},
  year = {2025}
}

@article{YuanFu2022,
  title = {Supercurrent diode effect and finite-momentum superconductivity in spin-orbit coupled systems},
  author = {Yuan, N. F. Q. and Fu, L.},
  journal = {Proc. Natl. Acad. Sci. USA},
  volume = {119},
  pages = {e2119548119},
  year = {2022}
}

@article{Bauriedl2022,
  title = {Supercurrent diode effect and magnetochiral anisotropy in crystalline NbSe$_2$},
  author = {Bauriedl, L. and B\"auml, C. and Fuchs, L. and Baumgartner, C. and Paulik, N. and Bauer, J. M. and Lin, K.-Q. and Lupton, J. M. and Taniguchi, T. and Watanabe, K. and Strunk, C. and Paradiso, N.},
  journal = {Nat. Commun.},
  volume = {13},
  pages = {4266},
  year = {2022}
}

@article{Strambini2022,
  title = {Field-free superconducting diode in van der Waals heterostructures},
  author = {Strambini, E. and Spies, M. and Ligato, N. and Ilic, S. and Rouco, M. and González-Orellana, C. and Ilyn, M. and Rogero, C. and Bergeret, F. S. and Moodera, J. S. and Virtanen, P. and Heikkilä, T. T. and Giazotto, F.},
  journal = {Nat. Commun.},
  volume = {13},
  pages = {2431},
  year = {2022}
}

@article{Yun2023,
  title = {Zero-field superconducting diode effect in topological materials},
  author = {Yun, J. and Son, S. and Shin, J. and Park, G. and Zhang, K. and Shin, Y. J. and Park, J.-G. and Kim, D.},
  journal = {Phys. Rev. Res.},
  volume = {5},
  pages = {L022064},
  year = {2023}
}

@article{Wu2022,
  title = {The field-free Josephson diode in a van der Waals heterostructure},
  author = {Wu, H. and Wang, Y. and Xu, Y. and Sivakumar, P. K. and Pasco, C. and Filippozzi, U. and Parkin, S. S. P. and Zeng, Y.-J. and McQueen, T. and Ali, M. N.},
  journal = {Nature},
  volume = {604},
  pages = {653},
  year = {2022}
}

@article{Sundaresh2023,
  title = {Field-free superconducting diode effect in hybrid Josephson junctions},
  author = {Sundaresh, A. and Väyrynen, J. I. and Lyanda-Geller, Y. and Rokhinson, L. P.},
  journal = {Nat. Commun.},
  volume = {14},
  pages = {1628},
  year = {2023}
}

@article{Hou2023,
  title = {Field-free superconducting diode},
  author = {Hou, Y. and Nichele, F. and Chi, H. and Lodesani, A. and Wu, Y. and Ritter, M. F. and Haxell, D. Z. and Davydova, M. and Ilic, S. and Glezakou-Elbert, O. and Varambally, A. and Bergeret, F. S. and Kamra, A. and Fu, L. and Lee, P. A. and Moodera, J. S.},
  journal = {Phys. Rev. Lett.},
  volume = {131},
  pages = {027001},
  year = {2023}
}

@article{Turini2022,
  title = {Gate-tunable superconducting diode effect in InSb nanoflag Josephson junctions},
  author = {Turini, B. and Salimian, S. and Carrega, M. and Iorio, A. and Strambini, E. and Giazotto, F. and Zannier, V. and Sorba, L. and Heun, S.},
  journal = {Nano Lett.},
  volume = {22},
  pages = {8502},
  year = {2022}
}

@article{Davydova2022,
  title = {Universal Josephson diode effect},
  author = {Davydova, M. and Prembabu, S. and Fu, L.},
  journal = {Sci. Adv.},
  volume = {8},
  pages = {eabo0309},
  year = {2022}
}

@article{Souto2022,
  title = {Supercurrent rectification from nonlocal tunneling in hybrid superconducting devices},
  author = {Souto, R. S. and Leijnse, M. and Schrade, C.},
  journal = {Phys. Rev. Lett.},
  volume = {129},
  pages = {267702},
  year = {2022}
}

@article{Wang2024,
  title = {Superconducting diode effect in Rashba spin–orbit coupled Josephson junctions},
  author = {Wang, J. and Jiang, Y. and Wang, J. J. and Liu, J.-F.},
  journal = {Phys. Rev. B},
  volume = {109},
  pages = {075412},
  year = {2024}
}

@article{Pillet2023,
  title = {Nonreciprocal Josephson effect in a ballistic quantum point contact},
  author = {Pillet, J.-D. and Annabi, S. and Peugeot, A. and Riechert, H. and Arrighi, E. and Griesmar, J. and Bretheau, L.},
  journal = {Phys. Rev. Res.},
  volume = {5},
  pages = {033199},
  year = {2023}
}

@article{Chieppa2025,
  title = {Unveiling the Current-Phase Relationship of InSb Nanoflag Josephson Junctions Using a NanoSQUID Magnetometer},
  author = {Chieppa, A. and Shukla, G. and Traverso, S. and Bucci, G. and Zannier, V. and Fracassi, S. and Ziani, N. T. and Sassetti, M. and Carrega, M. and Beltram, F. and Giazotto, F. and Sorba, L. and Heun, S.},
  journal = {Nano Lett.},
  volume = {25},
  pages = {14412},
  year = {2025}
}

@article{He2022,
  title = {Superconducting diode effect and anomalous Josephson current in quantum spin Hall edge junctions},
  author = {He, J. J. and Tanaka, Y. and Nagaosa, N.},
  journal = {New J. Phys.},
  volume = {24},
  pages = {053014},
  year = {2022}
}

@article{recon,
  title = {Spontaneous Breakdown of Topological Protection in Two Dimensions},
  author = {Wang, J. and Meir, Yigal and Gefen, Yuval},
  journal = {Phys. Rev. Lett.},
  volume = {118},
  pages = {046801},
  numpages = {5},
  year = {2017}
}

@article{Costa2023,
  title = {Theory of the superconducting diode effect in Rashba systems},
  author = {Costa, A. and Fabian, J. and Kochan, D.},
  journal = {Phys. Rev. B},
  volume = {108},
  pages = {054522},
  year = {2023}
}

@article{Coraiola2024,
  title = {Gate-tunable superconducting diode effect in planar Josephson junctions},
  author = {Coraiola, M. and Svetogorov, A. E. and Haxell, D. Z. and Sabonis, D. and Hinderling, M. and ten Kate, S. C. and Cheah, E. and Krizek, F. and Schott, R. and Wegscheider, W. and Cuevas, J. C. and Belzig, W. and Nichele, F.},
  journal = {ACS Nano},
  volume = {18},
  pages = {9221},
  year = {2024}
}

@article{dePicoli2023,
  title = {Superconducting diode effect in quasi-one-dimensional systems},
  author = {de Picoli, T. and Blood, Z. and Lyanda-Geller, Y. and Väyrynen, J. I.},
  journal = {Phys. Rev. B},
  volume = {107},
  pages = {224518},
  year = {2023}
}

@article{Hart2014,
  title = {Induced superconductivity in the quantum spin Hall edge},
  author = {Hart, S. and Ren, H. and Wagner, T. and Leubner, P. and Mühlbauer, M. and Brüne, C. and Buhmann, H. and Molenkamp, L. W. and Yacoby, A.},
  journal = {Nat. Phys.},
  volume = {10},
  pages = {638},
  year = {2014}
}

@article{Dolcini2015,
  title = {Topological Josephson $\phi_{0}$ junctions},
  author = {Dolcini, F. and Houzet, M. and Meyer, J. S.},
  journal = {Phys. Rev. B},
  volume = {92},
  pages = {035428},
  year = {2015}
}

@article{Wiedenmann2016,
  title = {4$\pi$-periodic Josephson supercurrent in HgTe-based topological Josephson junctions},
  author = {Wiedenmann, J. and Bocquillon, E. and Deacon, R. S. and Hartinger, S. and Herrmann, O. and Klapwijk, T. M. and Maier, L. and Ames, C. and Brüne, C. and Gould, C. and Oiwa, A. and Ishibashi, K. and Tarucha, S. and Buhmann, H. and Molenkamp, L. W.},
  journal = {Nat. Commun.},
  volume = {7},
  pages = {10303},
  year = {2016}
}

@article{Bernevig2006,
  title = {Quantum spin Hall effect and topological phase transition in HgTe quantum wells},
  author = {Bernevig, B. A. and Hughes, T. L. and Zhang, S.-C.},
  journal = {Science},
  volume = {314},
  pages = {1757},
  year = {2006}
}

@article{Konig2007,
  title = {Quantum spin Hall insulator state in HgTe quantum wells},
  author = {König, M. and Wiedmann, S. and Brüne, C. and Roth, A. and Buhmann, H. and Molenkamp, L. W. and Qi, X.-L. and Zhang, S.-C.},
  journal = {Science},
  volume = {318},
  pages = {766},
  year = {2007}
}

@article{Bernevig2006a,
  title = {Quantum Spin Hall Effect},
  author = {Bernevig, B. A. and Zhang, S.-C.},
  journal = {Phys. Rev. Lett.},
  volume = {96},
  pages = {106802},
  year = {2006}
}

@article{yin2020,
  title = {Tuning Rashba spin-orbit coupling at ${\mathrm{LaAlO}}_{3}/{\mathrm{SrTiO}}_{3}$ interfaces by band filling},
  author = {Yin, C. and Seiler, P. and Tang, L. M. K. and Leermakers, I. and Lebedev, N. and Zeitler, U. and Aarts, J.},
  journal = {Phys. Rev. B},
  volume = {101},
  pages = {245114},
  year = {2020}
}

@article{Zhi2025,
  title = {Tunable interfacial Rashba spin–orbit coupling in asymmetric Al$_x$In$_{1-x}$Sb/InSb/CdTe quantum well heterostructures},
  author = {Zhi, Z. and Wu, Y. and Ruan, H. and Liu, J. and Huang, P. and Yao, S. and Liu, X. and Tang, C. and Yao, Q. and Sun, L. and Zhang, Y. and Xiao, Y. and Che, R. and Kou, X.},
  journal = {Appl. Phys. Lett.},
  volume = {126},
  pages = {012104},
  year = {2025}
}

@article{Affandi2019,
  title = {Electric field-induced anisotropic Rashba splitting in two dimensional tungsten dichalcogenides WX$_2$ (X: S, Se, Te): A first-principles study},
  author = {Affandi, Y. and Absor, M. A. U.},
  journal = {Physica E},
  volume = {114},
  pages = {113611},
  year = {2019}
}

@article{Chen2021,
  title = {Tunable Rashba Spin Splitting in Two-Dimensional Polar Perovskites},
  author = {Chen, J. and Wu, K. and Hu, W. and Yang, J.},
  journal = {J. Phys. Chem. Lett.},
  volume = {12},
  pages = {1932--1939},
  year = {2021}
}

@article{Zhang2020,
  title = {Highly Efficient Electric-Field Control of Giant Rashba Spin–Orbit Coupling in Lattice-Matched InSb/CdTe Heterostructures},
  author = {Zhang, Y. and Xue, F. and Tang, C. and Li, J. and Liao, L. and Li, L. and Liu, X. and Yang, Y. and Song, C. and Kou, X.},
  journal = {ACS Nano},
  volume = {14},
  pages = {17396--17404},
  year = {2020}
}

@article{Schmidt2012,
  title = {Inelastic Electron Backscattering in a Generic Helical Edge Channel},
  author = {Schmidt, T. L. and Rachel, S. and von Oppen, F. and Glazman, L. I.},
  journal = {Phys. Rev. Lett.},
  volume = {108},
  pages = {156402},
  year = {2012}
}

@article{Ortiz2016,
  title = {Generic helical edge states due to Rashba spin-orbit coupling in a topological insulator},
  author = {Ortiz, L. and Molina, R. A. and Platero, G. and Lunde, A. M.},
  journal = {Phys. Rev. B},
  volume = {93},
  pages = {205431},
  year = {2016}
}

@article{Rothe2010,
  title = {Fingerprint of different spin-orbit terms for spin transport in HgTe quantum wells},
  author = {Rothe, D. G. and Reinthaler, R. W. and Liu, C.-X. and Molenkamp, L. W. and Zhang, S.-C. and Hankiewicz, E. M.},
  journal = {New J. Phys.},
  volume = {12},
  pages = {065012},
  year = {2010}
}

@article{Virtanen2012,
  title = {Signatures of Rashba spin-orbit interaction in the superconducting proximity effect in helical Luttinger liquids},
  author = {Virtanen, P. and Recher, P.},
  journal = {Phys. Rev. B},
  volume = {85},
  pages = {035310},
  year = {2012}
}

@article{Rothe2014,
  title = {Tunable polarization in a beam splitter based on two-dimensional topological insulators},
  author = {Rothe, D. G. and Hankiewicz, E. M.},
  journal = {Phys. Rev. B},
  volume = {89},
  pages = {035418},
  year = {2014}
}

@article{Liu2008,
  title = {Quantum Spin Hall Effect in Inverted Type-II Semiconductors},
  author = {Liu, C. and Hughes, T. L. and Qi, X.-L. and Wang, K. and Zhang, S.-C.},
  journal = {Phys. Rev. Lett.},
  volume = {100},
  pages = {236601},
  year = {2008}
}

@article{Konig2008,
  title = {The Quantum Spin Hall Effect: Theory and Experiment},
  author = {König, M. and Buhmann, H. and Molenkamp, L. W. and Hughes, T. L. and Liu, C.-X. and Qi, X.-L. and Zhang, S.-C.},
  journal = {J. Phys. Soc. Jpn.},
  volume = {77},
  pages = {031007},
  year = {2008}
}

@article{Ostrovsky2012,
  title = {Symmetries and weak localization and antilocalization of Dirac fermions in HgTe quantum wells},
  author = {Ostrovsky, P. M. and Gornyi, I. V. and Mirlin, A. D.},
  journal = {Phys. Rev. B},
  volume = {86},
  pages = {125323},
  year = {2012}
}

@article{Khouri2016,
  title = {High-temperature quantum Hall effect in finite gapped HgTe quantum wells},
  author = {Khouri, T. and Bendias, M. and Leubner, P. and Brüne, C. and Buhmann, H. and Molenkamp, L. W. and Zeitler, U. and Hussey, N. E. and Wiedmann, S.},
  journal = {Phys. Rev. B},
  volume = {93},
  pages = {125308},
  year = {2016}
}

@article{trenches,
  title = {Formation and detection of Majorana modes in quantum spin Hall trenches},
  author = {Fleckenstein, C. and Ziani, N. T. and Calzona, A. and Sassetti, M. and Trauzettel, B.},
  journal = {Phys. Rev. B},
  volume = {103},
  pages = {125303},
  year = {2021}
}

@incollection{Ginzburg1950,
  title = {On the Theory of Superconductivity},
  author = {Ginzburg, V. L. and Landau, L. D.},
  booktitle = {Men of Physics: L. D. Landau, Vol. 1, Low Temperature Physics},
  editor = {ter Haar, D.},
  publisher = {Pergamon Press},
  address = {Oxford},
  pages = {138--167},
  year = {1965}
}

@article{Fleckenstein2019,
  title = {Parafermions in Weakly Interacting Superconducting Constrictions at the Helical Edge of Quantum Spin Hall Insulators},
  author = {Fleckenstein, C. and Ziani, N. T. and Trauzettel, B.},
  journal = {Phys. Rev. Lett.},
  volume = {122},
  pages = {066801},
  year = {2019}
}

@article{Klinovaja2015,
  title = {Fractional charge and spin states in topological insulator constrictions},
  author = {Klinovaja, J. and Loss, D.},
  journal = {Phys. Rev. B},
  volume = {92},
  pages = {121410},
  year = {2015}
}

@article{Orth2015,
  title = {Non-Abelian parafermions in time-reversal-invariant interacting helical systems},
  author = {Orth, C. P. and Tiwari, R. P. and Meng, T. and Schmidt, T. L.},
  journal = {Phys. Rev. B},
  volume = {91},
  pages = {081406},
  year = {2015}
}

@article{Ronetti2020,
  title = {Levitons in helical liquids with Rashba spin-orbit coupling probed by a superconducting contact},
  author = {Ronetti, F. and Carrega, M. and Sassetti, M.},
  journal = {Phys. Rev. Res.},
  volume = {2},
  pages = {013203},
  year = {2020}
}

@article{Ghiasi2025,
  title = {Quantum spin Hall effect in magnetic graphene},
  author = {Ghiasi, T. S. and Petrosyan, D. and Ingla-Aynés, J. and Bras, T. and Watanabe, K. and Taniguchi, T. and Mañas-Valero, S. and Coronado, E. and Zollner, K. and Fabian, J. and Kim, P. and van der Zant, H. S. J.},
  journal = {Nat. Commun.},
  volume = {16},
  pages = {60377},
  year = {2025}
}

@article{Knez2011,
  title = {Evidence for Helical Edge Modes in InAs/GaSb Quantum Wells},
  author = {Knez, I. and Du, R.-R. and Sullivan, G.},
  journal = {Phys. Rev. Lett.},
  volume = {107},
  pages = {136603},
  year = {2011}
}

@article{Wu2018,
  title = {Observation of the Quantum Spin Hall Effect up to 100 Kelvin in a Monolayer Crystal},
  author = {Wu, S. and Fatemi, V. and Gibson, Q. D. and Watanabe, K. and Taniguchi, T. and Cava, R. J. and Jarillo-Herrero, P.},
  journal = {Science},
  volume = {359},
  pages = {76--79},
  year = {2018}
}

@article{Vigliotti2023,
  title = {Reconstruction-Induced $\varphi_{0}$ Josephson Effect in Quantum Spin Hall Constrictions},
  author = {Vigliotti, L. and Cavaliere, F. and Passetti, G. and Sassetti, M. and Ziani, N. T.},
  journal = {Nanomaterials},
  volume = {13},
  pages = {1497},
  year = {2023}
}

@article{Nadeem2023,
  title = {The superconducting diode effect},
  author = {Nadeem, M. and Fuhrer, M. S. and Wang, X.},
  journal = {Nat. Rev. Phys.},
  volume = {5},
  pages = {558},
  year = {2023}
}

@article{Bercioux2015,
  title = {Quantum transport in Rashba spin--orbit materials: a review},
  author = {Bercioux, D. and Lucignano, P.},
  journal = {Rep. Prog. Phys.},
  volume = {78},
  pages = {106001},
  year = {2015}
}

@article{Fu2008,
  title = {Superconducting Proximity Effect and Majorana Fermions at the Surface of a Topological Insulator},
  author = {Fu, L. and Kane, C. L.},
  journal = {Phys. Rev. Lett.},
  volume = {100},
  pages = {096407},
  year = {2008}
}

@article{Kitaev2001,
  title = {Unpaired Majorana fermions in quantum wires},
  author = {Kitaev, A. Yu.},
  journal = {Phys. Usp.},
  volume = {44},
  pages = {131},
  year = {2001}
}

@article{Fu2009,
  title = {Josephson current and noise at a superconductor/quantum-spin-Hall-insulator/superconductor junction},
  author = {Fu, L. and Kane, C. L.},
  journal = {Phys. Rev. B},
  volume = {79},
  pages = {161408},
  year = {2009}
}

@article{Bismuthene2017,
  title = {Bismuthene on a SiC substrate: A candidate for a high-temperature quantum spin Hall insulator},
  author = {Reis, F. and Li, G. and Dudy, L. and Bauernfeind, M. and Glass, S. and Hanke, W. and Thomale, R. and Schäfer, J. and Claessen, R.},
  journal = {Science},
  volume = {357},
  pages = {287},
  year = {2017}
}

@article{Xu2013,
  title = {Large-Gap Quantum Spin Hall Insulators in Tin Films},
  author = {Xu, Y. and Yan, B. and Zhang, H. and Wang, J. and Xu, G. and Tang, P. and Duan, W. and Zhang, S.-C.},
  journal = {Phys. Rev. Lett.},
  volume = {111},
  pages = {136804},
  year = {2013}
}

@article{Mueller2015,
  title = {Nonlocal transport via edge states in InAs/GaSb coupled quantum wells},
  author = {Mueller, S. and Pal, A. N. and Karalic, M. and Tschirky, T. and Charpentier, C. and Wegscheider, W. and Ensslin, K. and Ihn, T.},
  journal = {Phys. Rev. B},
  volume = {92},
  pages = {081303},
  year = {2015}
}

@article{Li2016,
  title = {Detection of Majorana Kramers Pairs Using a Quantum Point Contact},
  author = {Li, J. and Pan, W. and Bernevig, B. A. and Lutchyn, R. M.},
  journal = {Phys. Rev. Lett.},
  volume = {117},
  pages = {046804},
  year = {2016}
}

@article{Dolcetto2013,
  title = {Generating and controlling spin-polarized currents induced by a quantum spin Hall antidot},
  author = {Dolcetto, G. and Cavaliere, F. and Ferraro, D. and Sassetti, M.},
  journal = {Phys. Rev. B},
  volume = {87},
  pages = {085425},
  year = {2013}
}

@article{strunz,
  title = {Interacting topological edge channels},
  author = {Strunz, J. and Wiedenmann, J. and Fleckenstein, C. and Lunczer, L. and Beugeling, W. and Müller, V. L. and Shekhar, P. and Ziani, N. T. and Shamim, S. and Kleinlein, J. and Buhmann, H. and Trauzettel, B. and Molenkamp, L. W.},
  journal = {Nat. Phys.},
  volume = {16},
  pages = {83},
  year = {2020}
}

\end{document}


\title[Article Title]{\centering\textbf{Supplementary information}: Intrinsic and Tunable Superconducting Diode Effect in Quantum Spin Hall Systems}


\author*[1,2]{\fnm{Samuele} \sur{Fracassi}}\email{samuele.fracassi@edu.unige.it}

\author[1,2]{\fnm{Simone} \sur{Traverso}}

\author[3]{\fnm{Stefan} \sur{Heun}}

\author[1,2]{\fnm{Maura} \sur{Sassetti}}

\author[2]{\fnm{Matteo} \sur{Carrega}}

\author[1,2]{\fnm{Niccolo} \sur{Traverso Ziani}}

\affil[1]{\orgdiv{Dipartimento di Fisica}, \orgname{Università di Genova}, \orgaddress{\street{Via Dodecaneso 33}, \city{Genova}, \postcode{16146}, \state{Italy}}}

\affil[2]{\orgdiv{CNR-SPIN}, \orgaddress{\street{Via Dodecaneso 33}, \city{Genova}, \postcode{16146}, \state{Italy}}}

\affil[3]{\orgdiv{NEST}, \orgname{Istituto Nanoscienze-CNR and Scuola Normale Superiore}, \orgaddress{\street{Piazza San Silvestro 12}, \city{Pisa}, \postcode{56127}, \state{Italy}}}

\maketitle
\newpage
\section*{Beyond symmetry breaking for the presence or absence of SDE}
In the main text, we have presented the band structures responsible for supercurrent rectification. We now turn to highlight their common peculiar features, focusing on the microscopic mechanisms through which they are realized. While the breaking of spatial inversion and time-reversal symmetries is necessary for the occurrence of SDE, it is not sufficient. It should be recalled that understanding when this condition becomes sufficient is, at a general level, still an open problem. In this system, we have identified two specific requirements that mark the sufficiency of the condition: (i) an asymmetry of superconducting gaps between positive and negative $k$, and (ii) a $k$-dependent spin polarization of the excitations. When the symmetry breaking manifests at the band level through these features, rectification of the supercurrent is realized.\\ 
We now show how the situations discussed comply with this criterion.
We start with the wide structure geometry in the presence of inhomogeneous Rashba coupling $\theta_k \neq \chi_k$, discussed in Fig. 2 of the main text. One can note that fixing the energy of a single edge and reversing the sign of \(k\) corresponds to a flip in the spin polarity. For this reason, the presence of a magnetic field breaks the symmetry between $k$ and $-k$ on the single edge bands (if spin--orbit is homogeneous, the asymmetry vanishes in the total spectrum). Adding superconductivity leads hence to momentum-dependent gap openings. Correspondingly, the excitations have different total spin directions, and so the magnetic field lowers the superconducting gaps differently for positive and negative \(k\), as shown in Fig. 2\textbf{(b)} in the main text. This leads to a difference between opposite critical currents. The explanation of SDE for an inhomogeneous chemical potential shown in Fig. 3 is not very different from what we just discussed. The shift in chemical potential $\mu_u\neq \mu_d$ creates a difference between the absolute value of the momenta with opposite sign corresponding to the superconducting gaps, which implies that between the band minima the spin polarization is different, with a consequent different lowering of the gaps due to the magnetic field.\\
We now discuss the narrow well regime. We recall that when the symmetries are broken by an external magnetic field and spin-flip tunneling, the diode effect can only be observed if spin-preserving tunneling and spin–orbit coupling are also present, even if these do not break the symmetries any further. Indeed, without these additional requirements, the electronic fixed-energy states in absence of superconductivity have the form $|\textbf{n}_1,k\,\rangle$, $|\textbf{n}_2,-k\,\rangle$, $|\textbf{n}_3,k\,\rangle$, $|\textbf{n}_4,-k\,\rangle$, and the effect of symmetry breaking is projected onto the different $k$-independent directions $\textbf{n}_i$ of the total spin of the state. Since the fixed energy states appear in pairs with opposite momenta, particle-hole symmetry ensures that if $\Delta$ is large enough to make the system superconducting, the gaps open in a symmetric way in the $k$-space. This argument explicitly shows why, despite the necessary symmetry breakings, the mere presence of spin-flip tunneling and a magnetic field is not sufficient to obtain SDE. If we now consider the effect of spin-preserving tunneling, we find that the spin direction of the eigenstates acquires an explicit dependence on $k$. However, the energy crossing for the electron and hole sectors still occurs evenly in $k$. Taking into account the additional mixing due to spin--orbit coupling, the required superconducting gaps asymmetry manifests itself leading to a SDE through the mechanism described above.\\\\
Finally, we address the interesting case of finite SDE in the absence of any magnetic fields. Here, the symmetries are broken by spin-flip tunnelings unbalanced by edge reconstruction, which leads to supercurrent rectification if combined with the presence of spin--orbit coupling. Even without the latter, the superconducting gaps are asymmetrically opened in $k$ by reconstructed spin-flip tunneling. The missing ingredient, added by spin--orbit coupling, is the presence of a spin pattern of the eigenvectors that depends on $k$. With that included, the two requirements that lead to SDE are fulfilled.\\ Since here the SDE occurrence requires the spontaneous breaking of time-reversal symmetry, observing the intrinsic phenomenon in experiment in the regime where the edges are close enough to interact would represent a clear signature of the presence of edge reconstruction.\\
To further investigate the somehow uncommon case of intrinsic SDE, it is useful to start from Fig. 5 present in the main text, where one can identify regions in which the diode effect is absent despite the breaking of the two necessary symmetries. In particular, for specific values of the couple ($1/v_Fk_0,\mu$), one can observe $\eta=0$ independently of $b$.
To understand this, it is useful to observe that in order to destroy or enhance the SDE one should have, respectively, equal or maximally unequal spin-flip tunneling scattering amplitudes for right movers and left movers at the Fermi level in the gap closing moment. The spin-flip tunneling Hamiltonian $H_f$
in the basis of the well-defined direction of motion $\nu=\pm$, for which the kinetic energy is diagonal, takes the form
\begin{equation} H_f = \sum_k\tau_f(1+b\cos{2\theta_k})c^\dagger_{k+}d_{k+}-\tau_f(1-b\cos{2\theta_k})c^\dagger_{k-}d_{k-}+h.c. \end{equation}
Focusing on the situation where $\eta$ is suppressed, the momenta $\hat{k}$ with equal scattering amplitudes for opposite spin-flip tunneling satisfy
$ |\tau_f(1+b\cos{2\theta_{\hat{k}}})|=|-\tau_f(1-b\cos{2\theta_{\hat{k}}})|, $
which gives $\theta_{\hat{k}} = \pi/4+n\pi /2$, with $n \in \mathbb{N}$. In the approximation for $\theta_k$ that we used, $\hat{k}(n) = \pm\sqrt{(2n+1)\pi}k_0/2$. The next step is to link $\hat{k}$ and $\mu$. At gap closing, the momentum of gapless excitations depends on the details of the Hamiltonian. A qualitative estimate is given by $v_F|\hat{k}| \simeq \mu +\tau_f$.  These considerations lead to the family of curves 
\begin{equation}\frac{1}{v_Fk_0}\simeq \frac{\sqrt{(2n+1)\pi}}{2}\frac{1}{\mu+\tau_f}.
\label{iperb}\end{equation}
The presence of all other parameters creates a small deviation $\alpha(n)\equiv v_F|\hat{k}|-(\mu+\tau_f)$ for the gap closing momentum. In particular, the numerical evaluation of the bands shows that $\alpha(n=0)/\Delta \simeq 0.14$, $\alpha(n=1)/\Delta \simeq 0.04$.
This relation describes well the $\eta = 0$ regions in the parameter space where the symmetries necessary for SDE are broken. In Fig. 5\textbf{(a)} in the main text we show the $n=0$, $n=1$ hyperbolas defined by Eq. \ref{iperb} superimposed on the plot. At these specific points in the curves given by Eq. \ref{iperb}, the bands have superconducting gaps that are symmetric in $k$, and so for these specific parameters one of the two requirements for SDE is not satisfied. We underline that between the two hyperbolas with $\eta=0$ the rectification coefficient is optimized since in this region of the parameter space there is the curve that maximizes the imbalance of spin-flip tunneling processes due to edge reconstruction.
\section{Supplementary figures}

\begin{figure}[h]
    \centering
    \includegraphics[scale=0.5]{ 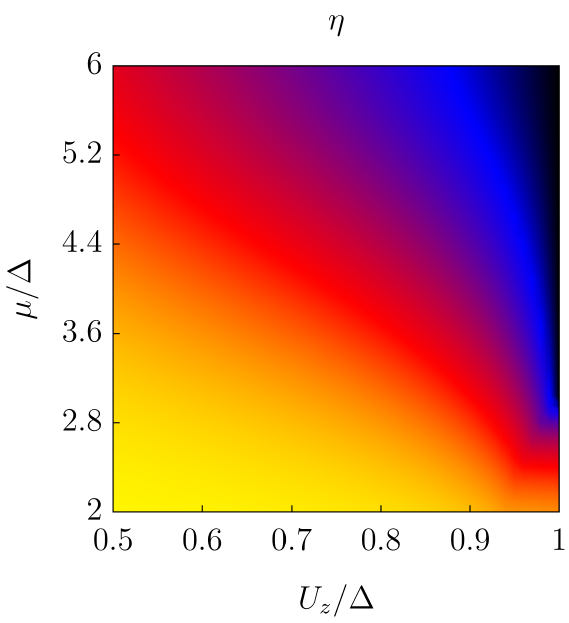} \includegraphics[scale=0.66]{ 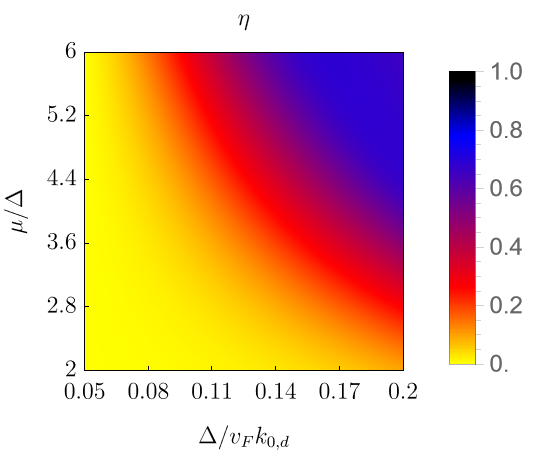}
    \caption{{\bf SDE performance with inhomogeneous spin--orbit coupling.} Rectification coefficient as a function of magnetic field $U_z/\Delta$, chemical potential $\mu/\Delta$, and Rashba inhomogeneity $\Delta/v_Fk_{0,d}$. In the two panels $\Delta/v_Fk_{0,u} = 0.05$. In the left panel $\Delta/v_Fk_{0,d} = 0.15$, in the right panel $U_z/\Delta = 0.8$.}
    \label{fig:placeholder1}
\end{figure}

\begin{figure}[h]
    \centering
    \includegraphics[scale=0.48]{ 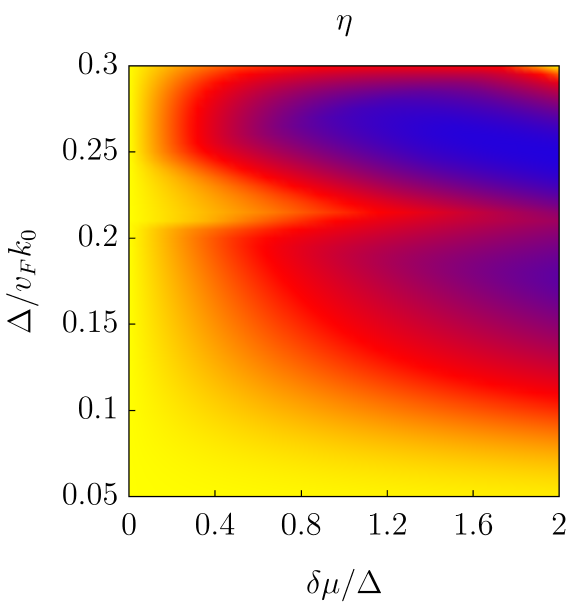} \includegraphics[scale=0.61]{ 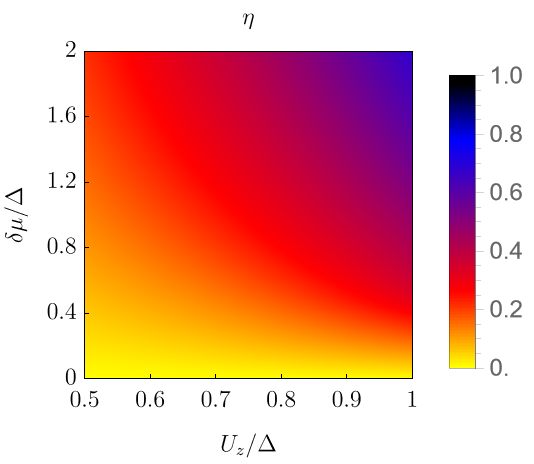}
    \caption{{\bf SDE performance with inhomogeneous chemical potentials.} Rectification coefficient as a function of magnetic  field $U_z/\Delta$, spin--orbit strength $1/v_Fk_0$, and chemical potential inhomogeneity $\delta \mu/\Delta$. In the left panel $U_z/\Delta = 0.85$, in the right panel $\Delta/v_Fk_0 = 0.2$. We fixed $(\mu_u+\mu_d)/2\Delta = 4$ in the two plots.}
    \label{fig:placeholder2}
\end{figure}

\begin{figure}[h]
    \centering
    \includegraphics[scale=0.62]{ 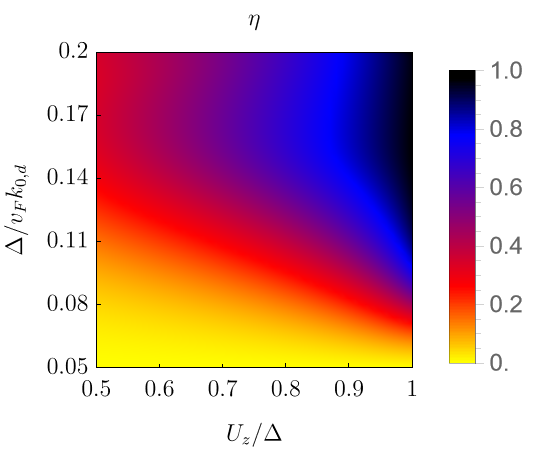} \includegraphics[scale=0.62]{ 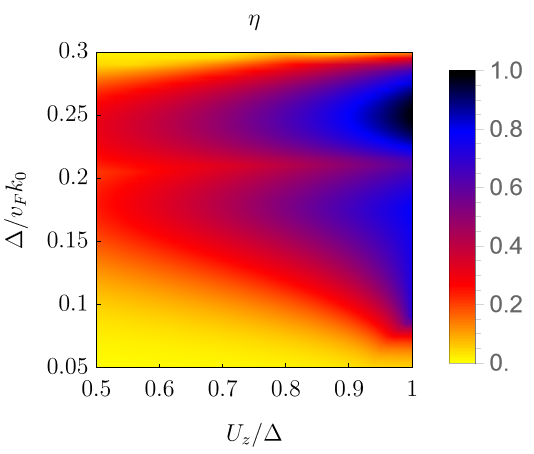}
    \caption{{\bf SDE performance with small tunneling processes.} In the left (inhomogeneous Rashba coupling) and right (inhomogeneous chemical potential) panel we can see that tunneling processes do not affect the physics described in the main text. Here we have set $\tau_f/\Delta = 0.1, \tau_p/\Delta=0.15$. The other parameters are the same as reported in Fig. 2\textbf{(a)} and Fig. 3\textbf{(a)} in the main text. }
    \label{fig:placeholder3}
\end{figure}

\begin{figure}[h]
    \centering
    
    \includegraphics[scale=0.42]{ 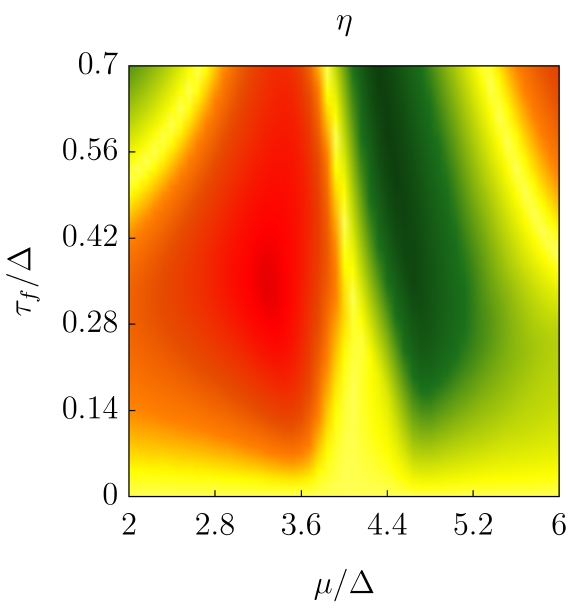} 
    \includegraphics[scale=0.54]{ 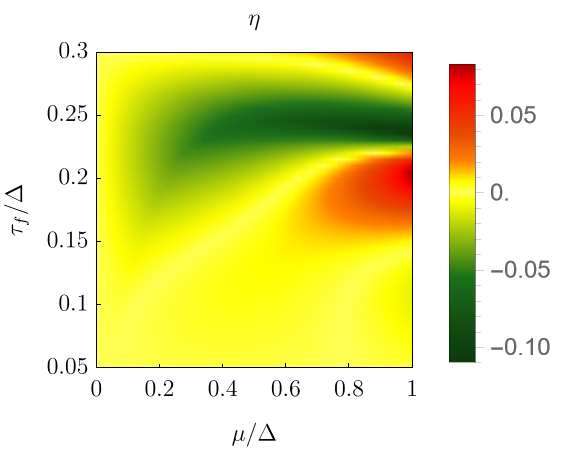}
    \caption{{\bf SDE without edge reconstruction. }Rectification coefficient as a function of magnetic field $U_z/\Delta$, spin--orbit strength $\Delta/v_Fk_0$, chemical potential $\mu/\Delta$, and spin-flipping tunneling energy scale $\tau_f/\Delta$.  In the left panel $U_z/\Delta = 0.85, 1/v_Fk_0 = 0.2$,
    in the right panel $\mu/\Delta = 4, \tau_f/\Delta = 0.6$. We fixed $\tau_p/\Delta = 0.7$ in these plots.}
    \label{fig:placeholder4}
\end{figure}

\begin{figure}[h]
    \centering
    \includegraphics[scale=0.42]{ 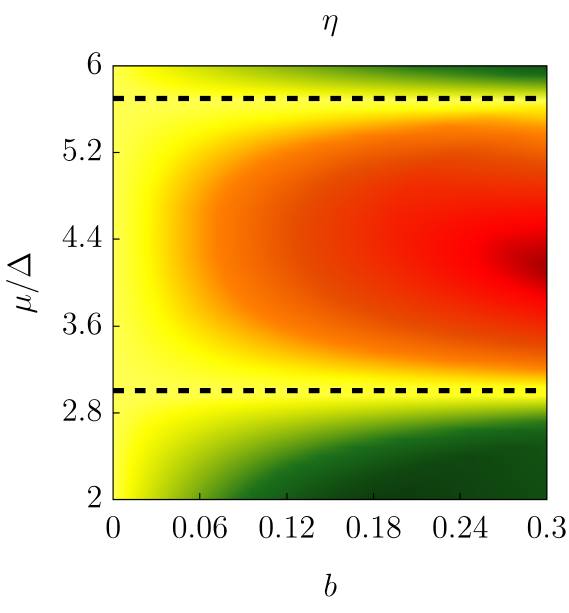} \includegraphics[scale=0.56]{ 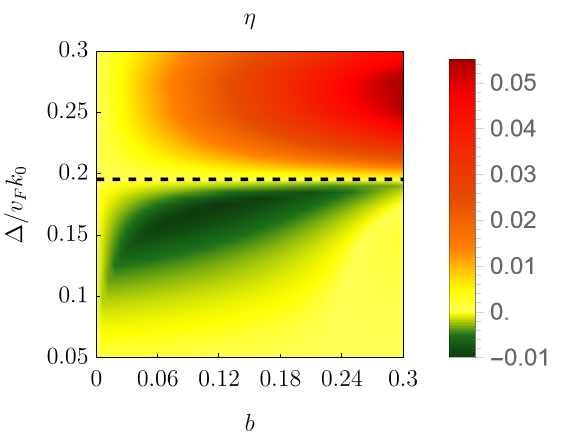}
    \caption{{\bf SDE with edge reconstruction ($b\neq0,a=0$). }Rectification coefficient as a function spin--orbit strength $\Delta/v_Fk_0$, chemical potential and spin-flip tunneling reconstruction coefficient $b$. Here we set $a=0$. In the left panel $\mu/\Delta = 4$, in the right panel $\Delta/v_Fk_0 = 0.2$. In the two plots $\tau_f/\Delta = 0.4, \tau_p/\Delta = 0.7$. Sharp lines of zero rectification are manifest at specific values of the couple $(\mu,1/v_Fk_0)$ that are independent of $b$. In the left panel we have $\eta=0$ for ($\mu/\Delta=3.005,\Delta/v_Fk_0 = 0.25$) that lie on the $n=0$ hyperbole defined by Eq. \ref{iperb} and for ($\mu/\Delta=5.699,\Delta/v_Fk_0 = 0.25$) that lie on the $n=1$ hyperbole. In the right panel we have $\eta=0$ for ($\mu/\Delta=4,\Delta/v_Fk_0 = 0.195$) that lie on the $n=0$ hyperbole.}
    \label{fig:placeholder5}
\end{figure}

\begin{figure}[h]
    \centering
    \includegraphics[scale=0.55]{ 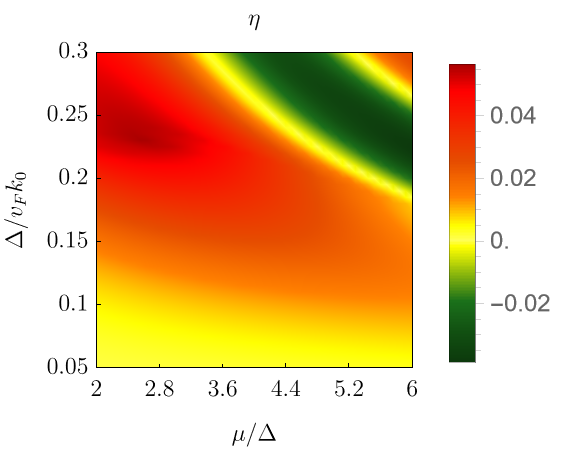}
    \caption{{\bf SDE with edge reconstruction ($b=0,a\neq0$).} Rectification coefficient as a function of spin--orbit strength $\Delta/v_Fk_0$ and chemical potential. Here $a=0.2, b=0$. We fixed the tunneling energy scales at $\tau_f/\Delta = 0.6, \tau_p/\Delta = 0.4$. Also here we can see two hyperbolas that give zero rectification for the mechanism described in the main text. This shows that also with $b=0, a\neq0$, SDE can manifest.}
    \label{fig:placeholder6}
\end{figure}

\begin{figure}[h]
    \centering
    \includegraphics[scale=0.55]{ 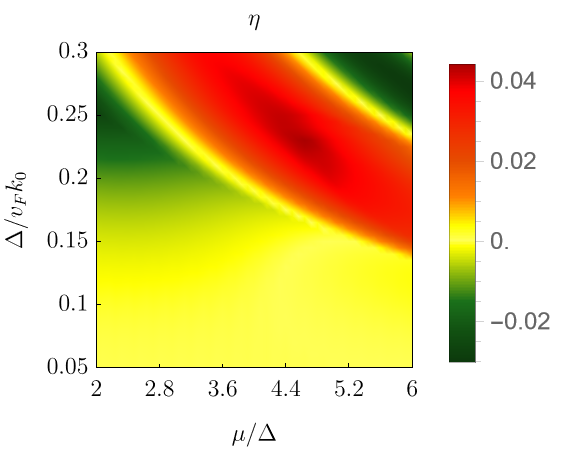}
    \caption{{\bf SDE with edge reconstruction ($b\neq0,a=0$) and without spin preserving tunneling.} Rectification coefficient as a function of spin--orbit strength $\Delta/v_Fk_0$ and chemical potential. Here $b=0.2, a=0$. We fixed the tunneling energy scales at $\tau_f/\Delta = 0.4, \tau_p/\Delta = 0$. This panel show that $\tau_p \neq0$ is not a necessary ingredient for the SDE in this system.}
    \label{fig:placeholder7}
\end{figure}

Here we show the dependence of $\eta$ on other parameters in the various regimes discussed in the main text. Moreover, we show additional plots that complete the arguments of the main text. In particular, Fig. \ref{fig:placeholder1} and Fig. \ref{fig:placeholder2} show the additional density plots of $\eta$ in the large regime. In Fig. \ref{fig:placeholder3} we show that small tunneling processes do not affect the physics discussed for large quantum wells. In  Fig. \ref{fig:placeholder4} and Fig. \ref{fig:placeholder5} show the additional density plots of $\eta$ in the narrow regime. In Fig. \ref{fig:placeholder6} we show the presence of $\eta \neq 0$ in the situation of reconstructed spin-preserving tunneling ($a \neq 0, b=0$). In Fig. \ref{fig:placeholder7} we show that spin preserving tunneling is not a necessary condition for the presence of $\eta \neq 0$ in the physical situation with reconstructed spin-flip tunneling.
\subsection{Large quantum wells}

\newpage

\newpage
\subsection{Narrow quantum wells}
